%% file: main.tex
\documentclass[journal]{IEEEtran}

\usepackage{booktabs} 

\usepackage{subfig}
\usepackage{pgf}
\usepackage{pgfplots,pgfplotstable} 
\usepgfplotslibrary{units}

\usepackage{epsfig}
\usepackage{xspace}
\usepackage{listings}
\usepackage{url}
\usepackage{pgf}
\usepackage{pgfplots,pgfplotstable} 
\usepgfplotslibrary{units}

\usepackage{mathrsfs}
\usepackage{filecontents}
\usepackage{tikz} 
\usetikzlibrary{arrows,shapes,automata,positioning,decorations,calc} 
\usetikzlibrary{spy,backgrounds,fit,shapes.misc}
\pgfdeclaredecoration{complete sines}{initial}
{
  \state{initial}[
  width=+0pt,
  next state=sine,
  persistent precomputation={\pgfmathsetmacro\matchinglength{
      \pgfdecoratedinputsegmentlength / int(\pgfdecoratedinputsegmentlength/\pgfdecorationsegmentlength)}
    \setlength{\pgfdecorationsegmentlength}{\matchinglength pt}
  }] {}
  \state{sine}[width=\pgfdecorationsegmentlength]{
    \pgfpathsine{\pgfpoint{0.25\pgfdecorationsegmentlength}{0.5\pgfdecorationsegmentamplitude}}
    \pgfpathcosine{\pgfpoint{0.25\pgfdecorationsegmentlength}{-0.5\pgfdecorationsegmentamplitude}}
    \pgfpathsine{\pgfpoint{0.25\pgfdecorationsegmentlength}{-0.5\pgfdecorationsegmentamplitude}}
    \pgfpathcosine{\pgfpoint{0.25\pgfdecorationsegmentlength}{0.5\pgfdecorationsegmentamplitude}}
  }
  \state{final}{}
}
\usepackage[latin1]{inputenc}

\usepackage[cmex10]{amsmath}
\usepackage{amsfonts,amssymb, amsthm} 

\usepackage[linesnumbered]{algorithm2e}
\SetCommentSty{mycommfont}
\SetAlFnt{\scriptsize}
\usepackage{array}
\usepackage{multirow}

\newtheorem{definition}{Definition}
\newtheorem{theorem}{Theorem}
\newtheorem{lemma}{Lemma}

\newtheorem{remark}{Remark}
\usepackage{enumitem}

\usepackage{xparse}
\usepackage{ifthen} 

\usepackage{acronym}
\usepackage[font=small,skip=0pt]{caption}

\makeatletter
\def\@copyrightspace{\relax}
\makeatother

\input{./00_macros.tex}  
\input{./00_acronyms.tex} 		
\begin{document}
\title{Quantized State Hybrid Automata for Cyber-Physical Systems}

\author{Avinash~Malik,
  Partha~Roop
  \thanks{A. Malik P. Roop are with the Department of Electrical and
    Computer Engineering, University of Auckland, Auckland, NZ e-mail:
    avinash.malik@auckland.ac.nz,
    p.roop@auckland.ac.nz.}
}

\maketitle

\begin{abstract}
  Cyber-physical systems involve a network of discrete controllers that
  control physical processes. Examples range from autonomous cars to
  implantable medical devices, which are highly safety critical.
  \acf{HA} based formal approach is gaining momentum for the
  specification and validation of CPS.\@ \ac{HA} combines the model of
  the plant along with its discrete controller resulting in a piece-wise
  continuous system with discontinuities. Accurate detection of these
  discontinuities, using appropriate level crossing detectors, is a key
  challenge to simulation of CPS based on \ac{HA}.

  Existing techniques employ time discrete numerical integration with
  bracketing for level crossing detection. These techniques 
  involve back-tracking and are highly non-deterministic and hence error
  prone. As level crossings happen based on the values of continuous
  variables, \acf{QSS}-integration may be more suitable. Existing \ac{QSS} integrators, based on fixed quanta, are also
  unsuitable for simulating \acp{HA}. This is since the quantum selected is not
  dependent on the \ac{HA} guard conditions, which are the main cause of  discontinuities. Considering
  this, we propose a new \textit{dynamic} quanta based formal model called
  \acf{QSHA}. The developed
  formal model and the associated simulation framework guarantees that (1)
  all level crossings are accurately detected and (2) the time of the
  level crossing is also accurate within floating point error bounds.
  Interestingly, benchmarking results reveal that the proposed
  simulation technique takes $720$, $1.33$ and $4.41$ times fewer
  simulation steps compared to standard \acf{QSS}-1, \acf{RK}-45, and
  \acf{DASSL} integration based techniques respectively.

\end{abstract}
\input{./intro}
\input{./prelim}
\input{./sands}
\input{./tech}

\input{./exp}
\input{./conclusion}


\begin{IEEEbiography}[{\includegraphics[width=1in,height=1.25in,clip,keepaspectratio]
    {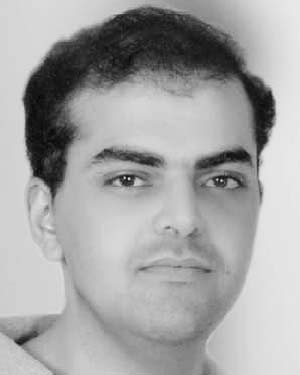}}]{Avinash Malik}
  is a senior lecturer at the University of Auckland, New Zealand. His
  main research interest lies in programming languages for multicore and
  distributed systems and their formal semantics and compilation. He has
  worked at organisations such as INRIA in France, Trinity College
  Dublin, IBM research Ireland, and IBM Watson on the design and the
  compilation of programming languages. He holds B.E. and Ph.D degrees
  from the University of Auckland.
\end{IEEEbiography}

\begin{IEEEbiography}[{\includegraphics[width=1in,height=1.25in,clip,keepaspectratio]
    {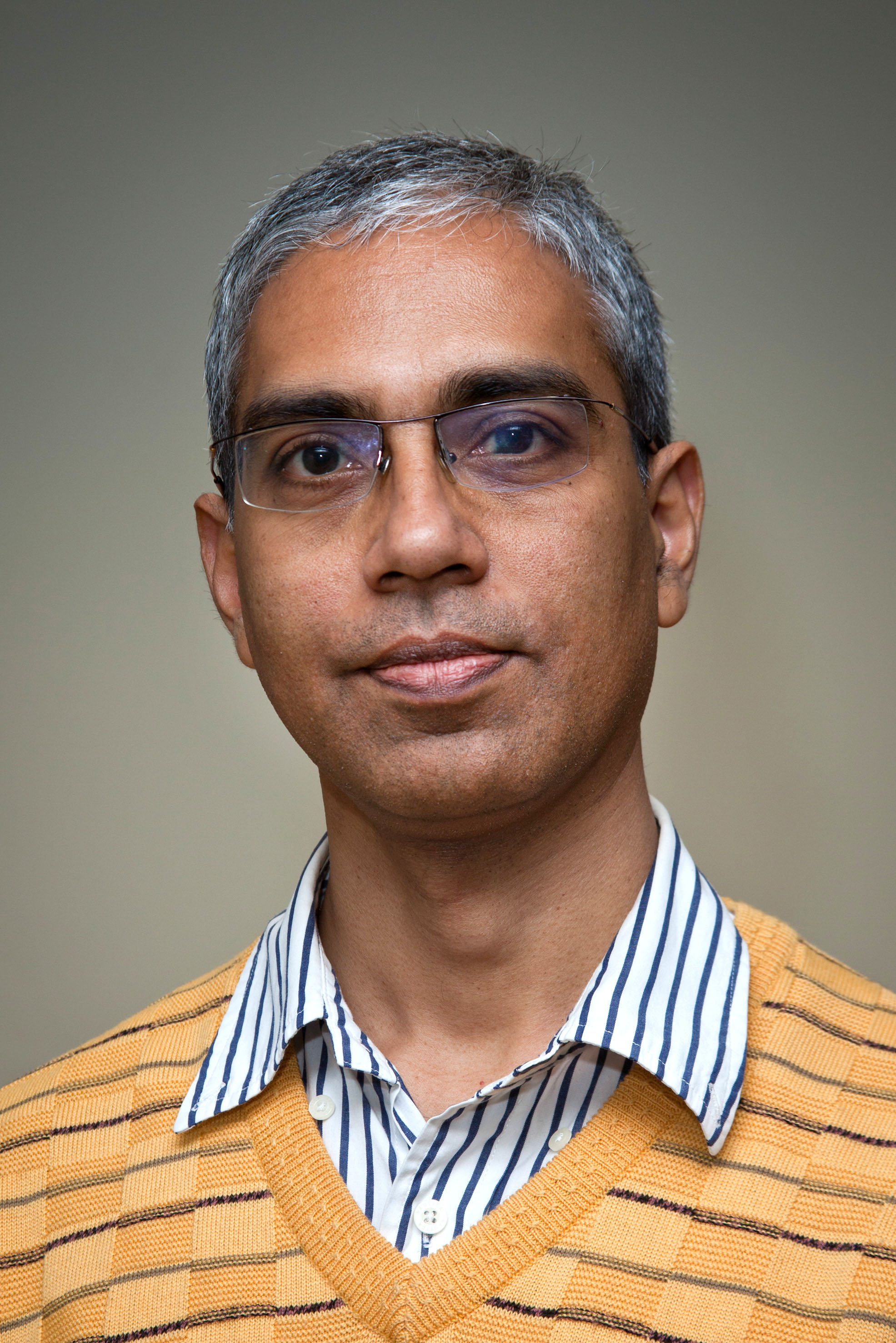}}]{Partha S. Roop}
  received his Ph.D degree in computer science (software engineering)
  from the University of New South Wales, Sydney, Australia, in 2001. He
  is currently an Associate Professor and is the Director of the
  Computer Systems Engineering Program with the Department of Electrical
  and Computer Engineering, the University of Auckland, New Zealand.
  Partha is an associated team member of the SPADES team INRIA,
  Rhone-Alpes, France, and held a visiting position in CAU Kiel,
  Germany, and Iowa State University, USA. His research interests
  include the design and verification of embedded systems. In
  particular, he is developing techniques for the design of embedded
  applications in automotive, robotics, and intelligent transportation
  systems that meet functional-safety standards.
\end{IEEEbiography}

\end{document}

%% file: 00_macros.tex
\newcommand{\ignore}[1]{{}}

\newcolumntype{L}[1]{>{\raggedright\let\newline\\\arraybackslash\hspace{0pt}}m{#1}}
\newcolumntype{C}[1]{>{\centering\let\newline\\\arraybackslash\hspace{0pt}}m{#1}}
\newcolumntype{R}[1]{>{\raggedleft\let\newline\\\arraybackslash\hspace{0pt}}m{#1}}



%% file: 00_acronyms.tex
\acrodef{CPS}{Cyber-physical Systems}
\acrodef{TA}{Timed Automata}
\acrodef{FSM}{Finite State Machine}
\acrodefplural{FSM}{Finite State Machine}
\acrodef{DES}{Discrete Event Simulation}
\acrodef{DSP}{Digital Signal Processor}
\acrodef{DTTS}{Discrete Time Transition System}
\acrodef{EA}{Evolutionary Algorithm}
\acrodef{HA}{Hybrid Automata}
\acrodef{HIOA}{Hybrid Input Output Automaton}
\acrodefplural{HIOA}{Hybrid Input Output Automata}
\acrodef{ILP}{Integer Linear Programming}
\acrodef{MCU}{Microcontroller Unit}
\acrodef{ODE}{Ordinary Differential Equation}
\acrodef{PoC}{Plant-on-a-Chip}
\acrodef{PLL}{Phase-locked loop}
\acrodef{QSS}{Quantized State System}
\acrodef{QSHA}{Quantized State Hybrid Automata}
\acrodef{SA}{Sinoatrial}
\acrodef{SHIOA}{Synchronous Hybrid Input Output  Automata}
\acrodef{SWIOA}{Synchronous Witness Input Output Automata}
\acrodef{TTS}{Time Triggered System}
\acrodef{WHIOA}{Well-formed Hybrid Input Output Automata}
\acrodef{WCET}{Worst-Case Execution Time}
\acrodef{WCRT}{Worst-Case Reaction Time}
\acrodef{SEA}{Synchronous Emulation Automaton}
\acrodefplural{SEA}{Synchronous Emulation Automata}
\acrodef{SWA}{Synchronous Witness Automaton}
\acrodefplural{SWA}{Synchronous Witness Automata}
\acrodef{DSL}{Domain Specific Language}
\acrodef{LTI}{Linear Time Invariant}
\acrodef{SMT}{Satisfiability Modulo Theory}
\acrodef{BMC}{Bounded Model Checking}
\acrodef{DHIOA}{Discrete Hybrid Input Output  Automata}
\acrodef{RHS}{Right Hand Side}
\acrodef{LHS}{Left Hand Side}
\acrodef{TH}{Thermostat}
\acrodef{WLM}{Water Level Monitor}
\acrodef{AFb}{Atrial Fibrillation}
\acrodef{Reactor}{Nuclear Reactor}
\acrodef{Robot}{Nonholonomic Robot}
\acrodef{RK}{Runge-Kutta}
\acrodef{DASSL}{Differential Algebraic System Solver}
\acrodef{DQSS}{Dynamic Quantized State System}

%% file: intro.tex
\section{Introduction}
\label{sec:introduction}

\begin{figure*}[tb]
  \centering

  \subfloat[The scenario that leads to missed level crossing
  detection\label{fig:robotnaviga}]{
    \includegraphics[scale=0.5]{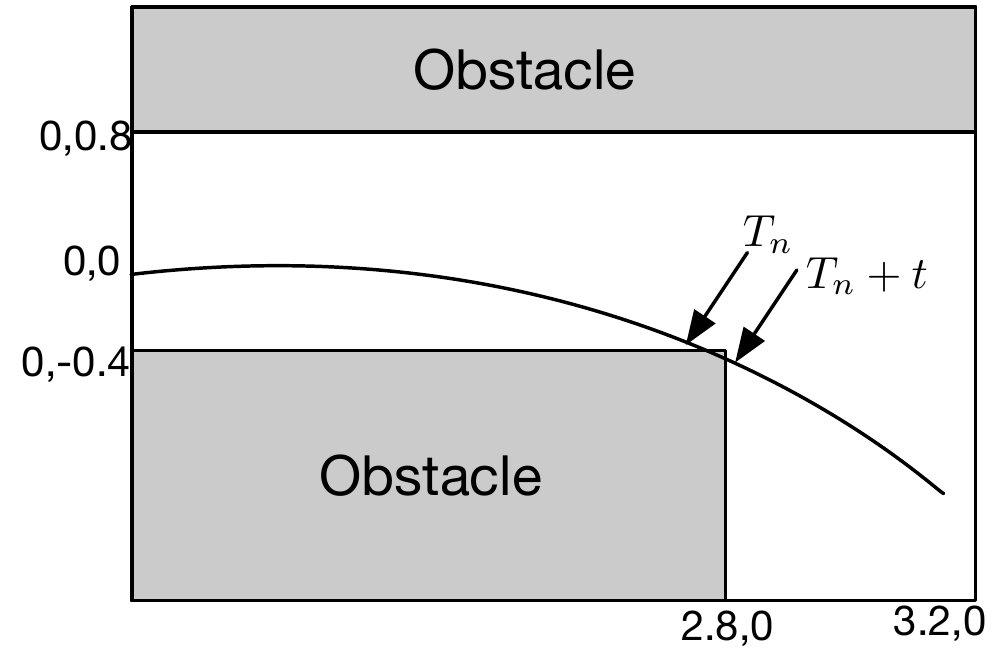}
  }
  \qquad
  \subfloat[The \ac{HA} describing the robot
  dynamics\label{fig:robotha}]{
    \scalebox{0.9}{\input{./robot.tex}}
  }
  \caption{Nonholonomic robot navigation in an obstacle course}
  \label{fig:robotnavig}
\end{figure*}

\ac{CPS} combine a set of discrete controllers, the cyber-part of the
system, which control a set of physical processes also known as the
plant. Such systems are highly safety critical in nature. Consider the
example of a nonholonomic robot trying to navigate in an obstacle ridden
environment as shown in Figure~\ref{fig:robotnaviga} and obtained
from~\cite{de1998feedback}. 

The solid line shows the trajectory that the robot takes. The robot is
moving using rear-wheel drive where $v_{1}$ and $v_{2}$ are the driving
and steering velocities respectively. These values are input by the
controller. The position of the robot in the Cartesian plane is denoted
by the continuous variables $x$ and $y$, the steering angle is denoted
by $\phi$, the orientation of the robot is given by $\theta$ and the
distance between the front and back wheel is given by the constant $l$.
The robot starts from position $(0, 0)$ and moves towards the end of the
room. It is required that the robot should not collide with any of the
obstacles. Any collision with an obstacle will lead to the failure of
the robot, which may have safety implications. Considering this,
\ac{CPS} simulation must have formal guarantees.

\ignore{
Moreover, the robot is moving using rear-wheel drive where $v_{1}$ and
$v_{2}$ are the driving and steering velocities, respectively, input by
the controller. The position of the robot in the Cartesian plane is
denoted by the continuous variables $x$ and $y$. The steering angle by
$\phi$, the orientation of the robot is given by $\theta$ and finally
the distance between the front and back wheel is given by the constant
$l$.}

\acf{HA} is a formal model, which is widely adopted in the \ac{CPS}
design flow.  Formal verification of \ac{CPS} is ideal for ensuring
safety. However, due to well known undecidability result regarding reachability~\cite{henzinger1998s}, such
verification is carried out using restricted subset of \ac{HA}, such
as the recent work using linear \ac{HA}~\cite{xie2017deriving}. This
limits the system being modelled and reduces applicability to
realistic problems. Consequently, in this paper, we focus on 
simulation of \ac{CPS} using \ac{HA}, so as to guarantee the fidelity
of such simulation.

Simulation of \ac{HA} is also non-trivial. 
Such an automaton
behaves piecewise continuously in every  \emph{mode}  until a sudden discontinuity happens. 
Discontinuities are induced when a controller switches the mode of a plant. 
The controller monitors the plant dynamics and
when the plant reaches a pre determined ``level'', the controller
switches the mode of the plant. This, in turn,  changes the behaviour of
the plant.

The \ac{HA} describing the closed-loop dynamics of the robot and the
controller is shown in Figure~\ref{fig:robotha}. The robot moves at a
constant steering and driving velocities in the obstacle free path, as
long as it does not collide with the obstacles. The change in the
continuous variables is captured in the location $\mathbf{T1}$ of the
\ac{HA}. As soon as the robot touches any of the obstacles, the robot
stops and changes direction as shown in location $\mathbf{T2}$ of the
\ac{HA}. Once the robot has steered clear of the obstacles, it goes back
to moving at the original steering and driving velocities, as shown by
the edge connecting location \textbf{T2} to \textbf{T1}. The collision
with boundaries of the shaded obstacles in Figure~\ref{fig:robotnaviga}
is captured by the guard condition $g$ in the \ac{HA}. The reset
relations $r$ highlight the discontinuous updates of the continuous
variables.



The most common technique, popularized by tools such as
Modelica~\cite{tiller2012introduction}, Simulink and
Stateflow~\cite{simulink1993mathworks,zhang2008zero}, for simulation of
hybrid systems (including \ac{HA}) is a two stage process:
\textcircled{1} numerical integration of \acp{ODE} followed by
\textcircled{2} level crossing detection. The procedure can be described
as follows:

\begin{enumerate}
\item From any given point in time $T_{n}$ (where $T_{n} = 0$ for the
  very first instant) take an integration step $t$.
\item Compute the value of continuous variables at time $T_{n}$ and
  $T_{n}+t$ without committing the integration step.
\item Evaluate the guard at time $T_{n}$ and time $T_{n}+t$, given the
  values of the continuous variables computed in point 2 above. If the
  guard evaluates to true its sign is considered to be positive.
  Otherwise, it is considered to be negative. \ignore{Calculate the
    \textit{sign} of the guard at time $T_{n}$ and $T_{n}+t$, given the
    values of the continuous variables computed in the previous step.}
  If the sign changes, a level crossing has occurred.
\item Upon detecting the level crossing, \textit{localization} is
  performed. Localization determines the most accurate time (within the
  bounds of floating point error) when the very first level crossing
  happened by performing a binary search between $T_{n}$ and $T_{n}+t$.
\item If there is a change in sign, a discrete mode switch is made at
  the detected time instant and the above steps are repeated.
\item If there is no change in sign, the integration step is committed,
  i.e., all the continuous variable values are advanced to step
  $T_{n}+t$ and the above steps are repeated.
\end{enumerate}

For the \ac{HA} in Figure~\ref{fig:robotha}, the simulation engine took
steps such that the robot is just before the obstacle at time $T_{n}$
and after integration of time step $t$, it will collide with the
obstacle at time $T_{n}+t$. The guard condition, determining the
collision, at time $T_{n}$ is \emph{false}, because predicate
$-0.4 \leq y \leq 0.8$ holds. After a single integration step of size
$t$ the guard condition remains \emph{false}, because the predicate
$3.2 \leq x \leq 2.8$ holds. Hence, it \textit{appears} that the guard
has not changed sign as the simulation engine fails to detect the level
crossing. This results in wrong behaviour of the \ac{HA}, which in turn
leads to a safety violation i.e. a collision with the obstacle. For
$t = 1.4e^{-4}$, $v1 = 30m/s$, and $v2 = -10m/s$,
OpenModelica~\cite{fritzson2006openmodelica} does not detect the
collision with obstacles for the example in Figure~\ref{fig:robotnavig}.


In the general case, any time a \ac{HA} crosses a level even number of
times, the simulation engine may miss the level crossing. Moreover, the
engine might detect some other level crossings rather than the first
one, which in turn results in incorrect localization. The primary issue
is that the step size $t$ of the integrator is chosen independent of the
guard conditions. Techniques have been proposed to make the selection of
step size sensitive to the guard
conditions~\cite{park1996state,esposito2001accurate}. However, these
approaches are too inefficient in practice, because they require
changing the classical numerical integration techniques by incorporating
Lie derivatives of the guard set. Furthermore, the most flexible
technique~\cite{esposito2001accurate} cannot handle cases when the
\acp{ODE} flow tangential to the guard set. To conclude, classical
level-crossing detection techniques are unable to correctly handle
discontinuities during simulation of network of \acp{HA}. 


Our major \textbf{contribution} in the paper is a quantized state
semantics and associated simulation framework for \acp{HA}, where sudden
discontinuities are correctly captured. Our technical contributions in
this paper can be listed as follows:

\begin{enumerate}
\item A new formal model, called \acf{QSHA}, and its semantics is
  proposed for the simulation of \ac{CPS}.
\item A dynamic quantum selection and discrete event simulation
  technique that converges to the first level crossing is developed.
This results in an efficient discrete event simulation engine for \ac{QSHA}.
\end{enumerate}

The rest of the paper is arranged as follows:
Section~\ref{sec:preliminaries} describes the preliminary details needed
to read the rest of the paper. Section~\ref{sec:syntax-semantics} gives
the formal syntax and quantized state semantics of \ac{HA} and
\ac{QSHA}. Section~\ref{sec:proposed-event-based} then goes on to
describe the discrete event simulation framework.
Section~\ref{sec:experimental-results} then compares the proposed
technique with the current state-of-the-art. We finally conclude in
Section~\ref{sec:conclusion}.


%% file: robot.tex
\begin{tikzpicture}[->,>=stealth',shorten >=1pt,auto,
node distance=5.4cm,
semithick,scale=0.7, transform shape]
\tikzstyle{every state}=[rectangle,rounded corners, minimum height =
1.2cm, text width=2.8cm, text centered,
fill=blue!20,draw=none,text=black, draw,line width=0.3mm]

\node[state,
label={[shift={(0,0.1)}]$\neg g$}, 
label={[shift={(1.5,0.1)}]\textbf{\texttt{T1}}}]
(T2)  {$\dot{x} = cos(\theta)\times v_{1}$ \\
  $\dot{y} = sin(\theta) \times v_{1}$\\
  $\dot{\theta} = tan(\phi/l) \times v_{1}$\\
  $\dot{\phi} = v_{2}$};

\node[state, 
label={[shift={(0,0.1)}]$ g$}, 
label={[rotate=0,shift={(-1.5,0.1)}]\textbf{\texttt{T2}}  }] 
(T3) [node distance=4.2cm, right of=T2]
{$\dot{x} = cos(\theta)\times v_{3}$ \\
  $\dot{y} = sin(\theta) \times v_{3}$\\
  $\dot{\theta} = tan(\phi/l) \times v_{3}$\\
  $\dot{\phi} = v_{4}$};

\draw[transform canvas={yshift=0.3em}] (T2) -- (T3) node [midway]
{$\frac{g}{r}$};

\draw[transform canvas={yshift=-0.3em}] (T3) -- (T2) node [midway]
{$\frac{\neg g}{r}$};

\draw[<-, dashed](T2.180) -- node[below] {initial} ++(-2cm,-0cm);
\draw[<-, dashed](T2.180) -- node[above] {
  $
  \begin{aligned}
    x(0) = y (0) = 0, \\
    \theta (0) = 0, \\
    \phi (0) = 1
  \end{aligned}
  $} ++(-2.5cm,-0cm);


\node[xshift=0.0cm, yshift=-1.4cm] (TABLE) 
at (T2)[text width=1.4cm] 
{
  \begin{tabular}{cl}
    $g:$ & {{$((y \geq 0.8) \vee (x \geq 3.2)) \vee ((y \leq -0.4) \wedge (x \leq 2.8))$}} \\
    $r:$ & $x' = x, y' = y, \theta' = \theta, \phi' = \phi$
  \end{tabular}
};

\end{tikzpicture}


%% file: prelim.tex
\section{Preliminaries --- quantized state integration with level
  crossing detection}
\label{sec:preliminaries}

\begin{figure}[b]
  \centering
  \subfloat[\ac{QSS}-1\label{fig:qss1}]{
    \includegraphics[scale=0.8]{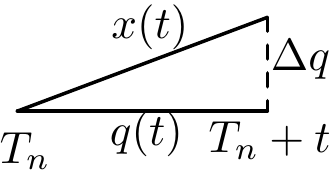}
  }
  \qquad
  \subfloat[\ac{QSS}-2\label{fig:qss2}]{
    \includegraphics[scale=0.8]{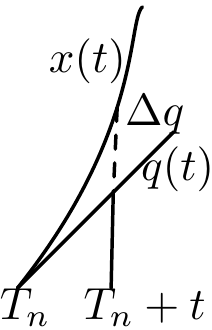}
  }
  \caption{\acf{QSS} techniques}
  \label{fig:qss}
\end{figure}

An integration technique that can be more amicable to hybrid system
simulation is that of quantizing the state (continuous variables) rather
than the time line. This technique is called
\acf{QSS}~\cite{kofman2004discrete,kofman2001quantized,kofman2002second,kofman2006third}.
\ac{QSS} techniques have been used to model \acp{HA} described in the
Modelica~\cite{tiller2012introduction} programming
language~\cite{floros2011automated,bergero2012simulating,di2016improving}.
\ac{QSS} based hybrid system simulators also follow the two stage
approach: \textcircled{1} integrate \acp{ODE} using \ac{QSS} techniques,
and \textcircled{2} perform level crossing detection. Herein we describe
the \ac{QSS} integration approaches, which will be necessary for reading
the rest of the paper.

Given a time-invariant \ac{ODE} system $ \dot{x} = f(x(t), t)$, like the
\acp{ODE} in the locations in Figure~\ref{fig:robotha}, where
$x(t) \in \mathbb{R}^{n}$ is the state vector. \ac{QSS} approximates the
\ac{ODE} as:
\begin{equation}
  \label{eq:qss}
  \dot{x} = f(q(t), t)
\end{equation}

\noindent
where $q(t)$ is the vector containing the quantized state variables,
which are quantized version of the state variables $x(t)$. Each variable
$q_{i}(t), i \in [1, n]$, can be approximated as a piecewise constant,
linear, quadratic or any high-order polynomial. We describe the two most
common approximations of $q_{i}(t)$, constant and linear called
\ac{QSS}-1 and \ac{QSS}-2, respectively.

\subsection{\acf{QSS}-1}
\label{sec:acqss-1}

The basic premise of first order \ac{QSS} called \ac{QSS}-1 is shown in
Figure~\ref{fig:qss1}. In case of \ac{QSS}-1, during integration, of any
\ac{ODE} $\dot{x}_{i}$, at any given point $T_{n}$ (for the very first
instant $T_{n} = 0$), a quantization function $q_{i}(t)$ is initialised
as $q_{i}(T_{n}) = x_{i}(T_{n})$. Next, $q_{i}(t)$ is held constant,
i.e., $q_{i}(t) = q_{i}(t^{-})$, by hysteresis, until some time $t$,
where $x_{i}(t)$ and $q_{i}(t)$ diverge by some user specified quantum
$\Delta q_{i}$. Formally, we solve a polynomial
$|x_{i}(t) - q_{i}(t)| = \Delta q_{i}$ to get the value of $t$, where
$\dot{x}_{i} = f(q_{1}(t),\ldots,q_{i}(t),\ldots,q_{n}(t), t)$.

Going back to our running example in Figure~\ref{fig:robotnavig}, for
each continuous variable $x$, $y$, $\theta$, $\phi$, in the system of
\acp{ODE} in location \textbf{T1} in Figure~\ref{fig:robotha}, we
initialize the functions $q_{x}(t)$, $q_{y}(t)$, $q_{\theta}(t)$, and
$q_{\phi}(t)$. From the initial conditions in the \ac{HA} we get
$q_{x}(0) = q_{y}(0) = q_{\theta}(0) = 0$ and $q_{\phi}(0) = 1$, for the
very first instant, $T_{n} = 0$. Similarly, we also have four quanta
$\Delta q_{x}$, $\Delta q_{y}$, $\Delta q_{\theta}$, and
$\Delta q_{\phi}$ for these continuous variables.

For continuous variable $x$, we have
$\dot{x} = f(q_{\theta}(t), t) = \cos(0) \times v_{1} = v_{1}$, because
$q_{\theta}(t) = q_{\theta}(t^{-}) = q_{\theta}(0) = 0$, by hysteresis. Hence, we solve
the polynomial \mbox{$|\int v_{1} d\tau - q_{x}(t)| = \Delta q_{x}$}. In this case,
it is simply \mbox{$|v_{1} \times t| = \Delta q_{x}$}, because
$q_{x}(t) = q_{x}(t^{-}) = q_{x}(0) = 0$ and $x(0) = 0$. In the general
case, the integral can be solved using simple explicit Euler technique.
Same for all other variables in our system of \acp{ODE}. \ac{QSS}-1
always results in a linear equation (first order polynomial) in $t$ to
be solved for its roots.

Once we have the value of variables at time $T_{n}$ ($T_{n} = 0$ for the
very first instant) and at time $T_{n}+t$, we can check for the change
in the sign of the \ac{HA} guard to detect level crossing using the
technique described in Section~\ref{sec:introduction}.

\subsection{\acf{QSS}-2}
\label{sec:acqss-2}

\ac{QSS}-2 differs from \ac{QSS}-1 in only the form of the quantized
variables. In case of \ac{QSS}-2, every quantized function $q_{i}(t)$ is
a linear equation, of the form:

\[
  q_{i}(t) = q_{i}(T_{n}) + \dot{x}_{i}(T_{n}) \times (t - T_{n})
\]

\noindent
For the very first instant, $T_{n} = 0$, we have
$q_{x}(0) = q_{y}(0) = q_{\theta}(0) = 0$ and $q_{\phi}(0) = 1$, like
before, for the running example. However, this time around, following
the linear approximation of quantized variables we get:
\[
  \begin{aligned}
    q_{x}(t) = 0 + \cos(\theta(0))\times v_{1} \times (t-0) = v_{1}\times t\\
    q_{y}(t) = 0 + \sin(\theta(0))\times v_{1} \times (t-0) = 0\\
    q_{\theta}(t) = 0 + \tan(\phi(0)/l)\times v_{1} \times (t-0)
    = \tan(l^{-1})\times v_{1}\times t\\
    q_{\phi}(t) = 0 + v_{2} \times (t-0) = v_{2}\times t
  \end{aligned}
\]

Next, considering, variable $x$, we have:
$\dot{x} = f(q_{\theta}(t), t) = \cos(\tan(l^{-1})\times v_{1}\times
t)$. Hence, we need to solve the equation:
\begin{equation}
  \label{eq:1}
  \bigl |
  \int \cos(\tan(l^{-1})\times v_{1}\times t) d\tau - q_{x}(t)
  \bigr | = \Delta q_{x}
\end{equation}
\noindent
to get $t$, where $x(t)$ and its quantized form $q_{x}(t)$ differ by
$\Delta q_{x}$, same for other variables in the system of \acp{ODE}.

\begin{gather}
  \cos(\tan(l^{-1})\times v_{1}\times t) =
  1 - \frac{(\tan(l^{-1})\times v_{1}\times t)^{2}}{2} \label{eq:c}\\
  \therefore \bigl |
  \int (1 - \frac{(\tan(l^{-1})\times v_{1}\times t)^{2}}{2})d\tau - q_{x}(t)
  \bigr | = \Delta q_{x} \label{eq:4}
\end{gather}

We can approximate the $\cos$ transcendental function by a Taylor
polynomial around zero, up to order one, as shown in
Equation~(\ref{eq:c}). We get a $n$-degree polynomial $x(t)$ after
integrating the Taylor polynomial (from Equation~(\ref{eq:4})), using
explicit forward Euler. Hence, \ac{QSS}-2 finds the time step $t$ where
the linear equation $q(t)$ and the $n$-degree polynomial equation $x(t)$
diverge by some user specified quanta as shown in Figure~\ref{fig:qss2}.

\ac{QSS} has been successfully used in formal frameworks for simulating
\ac{CPS} such as Ptolemy and
Modelica~\cite{LeeNiknamiNouiduiWetter15_ModelingSimulatingCyberPhysicalSystemsUsingCyPhySim,
  floros2011automated,bergero2012simulating}. It has also been
shown~\cite{kofman2004discrete} that \ac{QSS} is faster (in the number
of integration steps taken) for some \textit{stiff} systems compared to
classical numerical integration techniques like Runge-Kutta. However,
note that \ac{QSS} \textbf{like} classical numerical integration uses
the same level crossing detection technique. Hence, for any given
quantum can also miss level crossing events.

Furthermore, \ac{QSS}, unlike classical numerical integration, allows
integrators to execute asynchronously, which makes combination of
\ac{QSS} integration along with \ac{HA} ambiguous. Consider the guard
condition in the \ac{HA} in Figure~\ref{fig:robotha}. This guard
condition $g$ depends upon two variables $x$ and $y$, and hence when
evaluating the guard, at any point of time $T_{n}$ or $T_{n}+t$, value
of variables $x$ and $y$ should be available. However, since \ac{QSS}
integrators computing the values of these variables run asynchronously,
the values of $x$ and $y$ may \textit{not} be available together at any
time instant and hence, the guard cannot be evaluated, leading to
incorrect behaviour. Such semantic ambiguities need to be carefully
considered.


%% file: sands.tex
\section{\acf{QSHA} -- syntax and semantics}
\label{sec:syntax-semantics}

This section formalises the syntax and semantics of \ac{QSHA}, which
uses a new \ac{QSS} integration technique with \ac{HA}. We start by
defining \acf{HA} using Definition~\ref{def:ha}. When the model
comprises of multiple \acp{HA}, then the product \ac{HA} is obtained
using an asynchronous cross-product in the standard way~\cite{alur93}.

Consider the nonholonomic robot \ac{HA} shown in
Figure~\ref{fig:robotha} for illustration. This \ac{HA} has two
locations and hence $\mathbf{L}=\{\textbf{T1}, \textbf{T2}\}$. The
continuous variables used in the model are $x$, $y$, $\phi$, $\theta$ i.e.,
$X = \{x, y, \phi, \theta\}$ and $\mathbf{X} = \mathbb{R}^{4}$. Location
\textbf{T1} marked as the initial (or starting) location, hence
$Init = \{\mathbf{T1}\} \times \{0\} \times \{0\} \times \{0\} \times \{1\}$. The \acp{ODE},
inside the locations, capture the evolution of these continuous
variables. Formally, \acp{ODE} are represented as vector fields, e.g.,
$f(\mathbf{T1}, x, t, \phi, \theta)\ {\overset {\underset {\mathrm {def}
    }{}}{=}} \ {[\dot{x}, \dot{y}, \dot{\phi}, \dot{\theta}]}^{T} = {[\cos(\theta) \times
  v_1, \sin(\theta) \times v_1, \tan(\theta/l) \times v_1,
  v_2]}^{T}$\footnote{$^{T}$ indicates transpose of a matrix/vector.}.
Invariants are used as \emph{fairness conditions} to enable an exit from
any location as soon as the invariants become false. They also restrict
all continuous variables to obey the invariant conditions, while the
execution remains in a given location i.e. the invariant for the
location \textbf{T2} is
$g=((y \geq 0.8) \vee (x \geq 3.2)) \vee ((y \leq -0.4) \wedge (x \leq 2.8))$. There are two
edges $e_{1} = (\mathbf{T1}, \mathbf{T2})$,
$e_{2} = (\mathbf{T2}, \mathbf{T1})$ in $E$. An example guard on edge
$e_{1}$ is given by $G(e_{1}) = g$, which is also the invariant in
location $\mathbf{T2}$. Similarly, the reset relation on edge $e_{1}$ is
given as
$R(e_{1}, x, y, \theta, \phi)\ {\overset {\underset {\mathrm {def} }{}}{=}}\
{[x', y', \theta', \phi']}^{T} := {[x, y, \theta, \phi]}^{T}$, where
$x'$, $y'$, $\theta'$, and $\phi'$ give the updated values of the continuous
variables.

\begin{definition}
\label{def:ha}
A \acf{HA} is
$\mathcal{H} = \langle  L, X, Init, f, h, Inv, E, G, R
\rangle$, where:
  \begin{itemize}
    \setlength\itemsep{1pt}
  \item $\mathbf{L}$ a set of discrete locations.
  \item $X$ is a finite collection of continuous variables, with its
    domain represented as $\mathbf{X} = \mathbb{R}^n$. 
  \item $Init \subseteq \{l_0\} \times \mathbf{X}$
    such that there is exactly one $l_0 \in \mathbf{L}$, is the
    singleton initial location.
  \item
    $f : \mathbf{L} \times \mathbf{X} \rightarrow \mathbb{R}^{n}$
     is a vector field. 
  \item $Inv: \mathbf{L} \rightarrow 2^{\mathbf{X} }$ assigns
    to each $l \in \mathbf{L}$ an invariant set.
  \item $E \subset \mathbf{L} \times \mathbf{L} $ is a collection of discrete edges.
  \item $G : E \rightarrow 2^{\mathbf{X}}$ assigns to
    each $e = (l, l') \in E$ a guard.
  \item $R : E \times \mathbf{X} \rightarrow \mathbf{X}$ assigns to each
    $e = (l,l') \in E$, $x \in \mathbf{X}$ a reset relation.
  \end{itemize}
\end{definition}


\begin{definition}
  \label{def:guards}
  The edge guards $G$ are of the form:

  $g {\overset {\underset {\mathrm {def} }{}}{=}} x \bowtie \mathbb{Q} |
  g \wedge g | g \vee g$
\end{definition}
\noindent
where $x \in X$ is a continuous variable,
$\bowtie\ \in \{\geq, \leq, >, <\}$, and negation is expressed by reversing the
operator in $\bowtie$.

\begin{definition}
  For a \ac{HA}
  $\mathcal{H} = \langle L, X, Init, f, h, Inv, E, G, R\rangle$. For some outgoing
  edge guard $g \in G$, for some edge $e \in E$, let
  $g = p_{1} \wedge p_{2}$, where
  $p_{1} {\overset {\underset {\mathrm {def} }{}}{=}}\ l_{x}\ \tilde{\bowtie}\
  x\ \tilde{\bowtie}\ u_{x}$ and
  $p_{2} {\overset {\underset {\mathrm {def} }{}}{=}}\ l_{y}\ \tilde{\bowtie}\
  y\ \tilde{\bowtie}\ u_{y}$, where $\tilde{\bowtie} \in \{<, \leq\}$ and
  $x, y \in X$ and $l_{x}, u_{x}, l_{y}, u_{y} \in \mathbb{Q}$. Furthermore,
  there exists some
  $t^{l}_{x}, t^{u}_{x}, t^{l}_{y}, t^{u}_{y} \in \mathbb{R}$, such that
  $x(t^{l}_{x})-l_{x} = 0$, $x(t^{u}_{x})-u_{x}=0$,
  $y(t^{l}_{y})-l_{y}=0$, $y(t^{u}_{y})-u_{y}=0$ and
  $\forall t' < t^{l}_{x}, x(t^{l}_{x})-l_{x} \neq 0$,
  $\forall t' < t^{u}_{x}, x(t^{u}_{x})-u_{x}\neq0$,
  $\forall t' < t^{l}_{y}, y(t^{l}_{y})-l_{y}\neq0$,
  $\forall t' < t^{u}_{y}, y(t^{u}_{y})-u_{y}\neq0$. Then, $\mathcal{H}$ is
  \textit{well-formed} \textrm{iff}
  $t^{l}_{x} < t^{l}_{y} \leq t^{u}_{x}$ $\bigvee$
  $t^{l}_{y} < t^{l}_{x} \leq t^{u}_{y}$ $\bigvee t^{l}_{x} = t^{l}_{y}$
\end{definition}

\ac{QSHA} is defined using Definition~\ref{def:qsha}.

\begin{definition}
\label{def:qsha}

A \acf{QSHA} is
$\mathcal{H}_{q} = \langle L, X, X_q, Init, f, f_q, h, Inv, E, G, R
\rangle$, corresponding to a given \ac{HA}
$\mathcal{H} = \langle L, X, Init, f, h, Inv, E, G, R \rangle$. Here for
every variable $x \in X$, we introduce a quantised state variable
$q \in X_q$. The quantized state version of the vector field $f_q$ is
obtained from $f$ as follows: Given any $ \dot{x} = f(x(t), t)$ , an
\ac{ODE} defined using $f$, we introduce its quantized state \ac{ODE} in
$f_q$ as follows: $\dot{x} = f(q(t), t)$.
\end{definition}


The semantics of \ac{QSHA} is based on the concept of \emph{hybrid
  timesets with dynamic quanta}, hereafter referred to as hybrid
timesets, formalised using Definition~\ref{def:timeset}. This semantics
uses a dynamic quantum based integration step.

\begin{figure}[thb]
  \centering \includegraphics[scale=0.35]{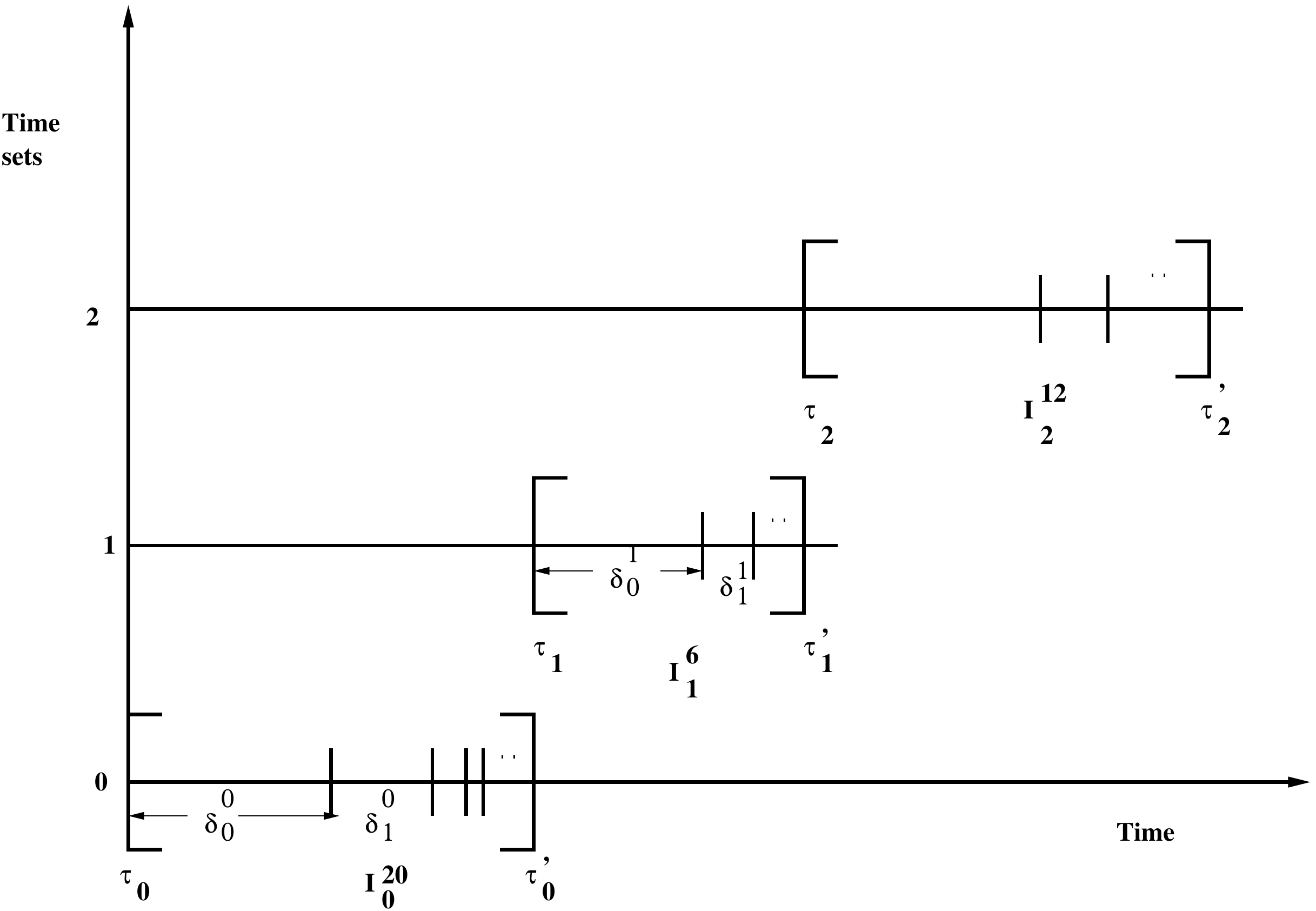}
  \caption{\ac{QSHA} semantics}
  \label{fig:timeset}
\end{figure}

Figure~\ref{fig:timeset} visualises the proposed semantics. Here, time
progresses in any \mbox{location} in discrete steps, which is captured
as a time interval $I_i^{k_i}$. During this time interval, the \ac{QSHA}
takes $k_i$ discrete steps, where each step corresponds to an
integration step of length $\delta_{0}$..$\delta_{k_i}$. The duration of each of
these integration steps is computed using the algorithm in
Section~\ref{sec:proposed-event-based}. Each of these $k_i$-steps are
known as \emph{intra-location} transitions, as control resides in a
given location. Exactly after the passage of $k_i$-th integration step,
an \emph{inter-location} transition is enabled to facilitate an
instantaneous \mbox{location} switch. This inter-location transition
takes zero time. In Figure~\ref{fig:timeset}, three intervals
$I_0^{20}, I_1^{6}$ and $I_2^{12}$ are visualised. The first interval
has 20 sub-intervals of varying lengths indicated by $\delta_0^0$ to
$\delta_{20}^0$, where the time intervals gradually become shorter. Each time
interval indicates the duration of an integration step. Informally, the
gradually diminishing time intervals make sense, because we are
approaching the inter-location transition. A formal reason is provided
later in Theorem~\ref{thm:2}, Section~\ref{sec:prop-prop-dynam}.
Likewise, the next two intervals have 6 and 12, gradually diminishing,
integration steps.

\begin{definition}
\label{def:timeset}
{\color{black}
A discrete hybrid timeset is a sequence of intervals $\tau = \{I_0^{k_0}, 
\ldots, I_{i}^{k_{i}}, \ldots, I_n^{k_n} \}$, such that:

\begin{enumerate}
  \setlength\itemsep{0pt}
\item
  $I_i^{k_i} = [ \tau_i=\tau_i^0, \tau_i^1=\tau_i + \delta_{0}^i, ...,
  \tau_i'=\tau_i + k_i \times \delta_{k_i}^i]$
\item If $n < \infty$ then
  $I_n^{k_n} = [ \tau_n=\tau_n^0, \tau_n^1=\tau_n + \delta_{0}^n, ...,
  \tau_n'=\tau_n + k_n \times \delta_{k_n}^n]$, or
  $I_n^{k_n} = [ \tau_n=\tau_n^0, \tau_n^1=\tau_n + \delta_{0}^n,
  ...,
  \tau_n'=\tau_n + k_n \times \delta_{k_n}^n)$, and
\item $\tau_i \leq \tau_i'$ and $\tau_i'  = \tau_{i+1}$ for all
  $i < n$.
\end{enumerate}}
\end{definition}

\begin{remark}
 Given $\tau = \{I_0^{k_0}, I_1^{k_1}, ..., I_n^{k_n} \}$
 \begin{enumerate}
   \setlength\itemsep{0pt}
 \item {\color{black} We denote {\boldmath$\tau$} as the domain of
     $\tau$, which includes any time instant $t$ that is in any interval
     $I_i^{k_i} \in \tau$.}
 \item Each interval $I_i^{k_i}$ has $k_i$ sub-intervals where
   $(\tau_i'-\tau_i)= \Sigma_{j=0}^{k_i}\delta_j^i$, and
 \item For any two consecutive intervals, $I_i^{k_i}$ and
   $I_{i+1}^{k_{i+1}}$, the ending time of the first interval equals the
   start time of the second interval i.e. $(\tau_i' = \tau_{i+1})$.
 \item The instantaneous separation between the time intervals
   $I_i^{k_i}$ and $I_{i+1}^{k_i+1}$ enables instantaneous mode
   switches. This in turn models infinite state changes in finite time,
   thereby modelling Zeno artefacts.
 \end{enumerate}
\end{remark}

The semantics of \ac{QSHA} is specified as a set of\emph{executions}. A
given execution is defined over a hybrid timeset $\tau$. The execution
over $\tau$ must satisfy the following three conditions: \textcircled{1}
any execution begins in an initial location. \textcircled{2} While
executing in a location, a set of intra-location transitions are taken
where time progresses in steps of variable length $\delta$. Starting at
any given time $T_n$ continuous variables evolve based on the specified
\acp{ODE} using the variable quantum based integration technique.
Moreover, the location invariant must hold. \textcircled{3} The
inter-location transitions are taken when the guard is satisfied and as
the transition triggers, some continuous variables are reset according
to specified reset relations. This is formalised in
Definition~\ref{def:execution}.

\begin{definition}
\label{def:execution}
Let $\Gamma$ be a collection of hybrid timesets.
A discrete execution of an \ac{QSHA} $\mathcal{H}_{q}$ is a three tuple
${\mathcal X}=(\tau, l, x)$ with $\tau \in \Gamma$,
$l:$ {\boldmath$\tau$} $\rightarrow \mathbf{L}$,
$x:$  {\boldmath$\tau$} $\rightarrow \mathbf{X}$,
satisfying the following:

\begin{enumerate}
  \setlength\itemsep{0pt}
\item Initial Condition: $(l(\tau_0), x(\tau_0)) \in Init$
\item Intra location transitions: For any interval
  $I_i^{k_i}=[\tau_i, \tau_i'] \in \tau$ the following must hold:
\begin{itemize}
\item Invariant satisfaction: for all
  $t \in \{ \tau_i, \tau_i+ \delta_0^i, \ldots, \tau_i+k_i \times
  \delta_{k_i}^i\}$, $x(t) \in Inv(l(t))$.
\item Continuous variable updates: for all $0 \leq j \leq k_i$

  $x(\tau_i) = x(\tau_i^0)$ and \\
  $x(\tau_i^{j+1}) = x(\tau_i^j) + \Delta_x(\delta_j^i)$, where
  $\Delta_x(\delta_j^i)$ is a function that selects an appropriate
  quantum $\Delta_{x_m}$ for the continuous variable $x_m$, where
  $1\leq m \leq n$, corresponding to the current integration step of
  length $\delta_j^i$.
\end{itemize}
\item Inter-location transitions: For all $i$, $e_i = (l(\tau_i'),
  l(\tau_{i+1})) \in E$, $x(\tau_i') \in G(e_{i})$ and
  $(x(\tau_{i+1}) \in R(e_i, x(\tau_i'))$.
\end{enumerate}
\end{definition}


%% file: tech.tex
\section{Discrete event simulation of \acf{QSHA}}
\label{sec:proposed-event-based}

Section~\ref{sec:syntax-semantics} fulfils the first objective of the
paper, by providing a formal quantized state semantics of \ac{QSHA}.
This section details the discrete event simulation technique, which
guarantees that the generated trace adheres to the semantics.

The semantics of \ac{QSHA}, represented as executions
(c.f.~Definition~\ref{def:execution}), captures a set of timed traces of
the automaton. Every execution must begin in the initial location $l_0$.
Execution then progresses from one location to the next when an
\emph{inter-location} transition is taken instantaneously following
Definition~\ref{def:execution}, rule 3. When execution begins in a new
location, the values of the continuous variables are updated using a set
of $\delta_0^i .. \delta_{k_i}^i$ steps of gradually diminishing integration steps
in the interval $I_i^{k_i}$. These are known as intra-location
transitions following Definition~\ref{def:execution}, rule 2. The
duration of integration steps are decided using a new \ac{QSS} algorithm
based on dynamic quanta, we term \acs{DQSS}, which is elaborated next.

\subsection{Computing the next scheduling event for each continuous
  variable}
\label{sec:comp-next-sched}

\begin{figure}[tb]
  \centering
  \subfloat[The primary idea for finding the next time event using a
  dynamic quanta obtained from the level
  crossing\label{fig:nextEventd}]{
    \includegraphics[scale=0.55]{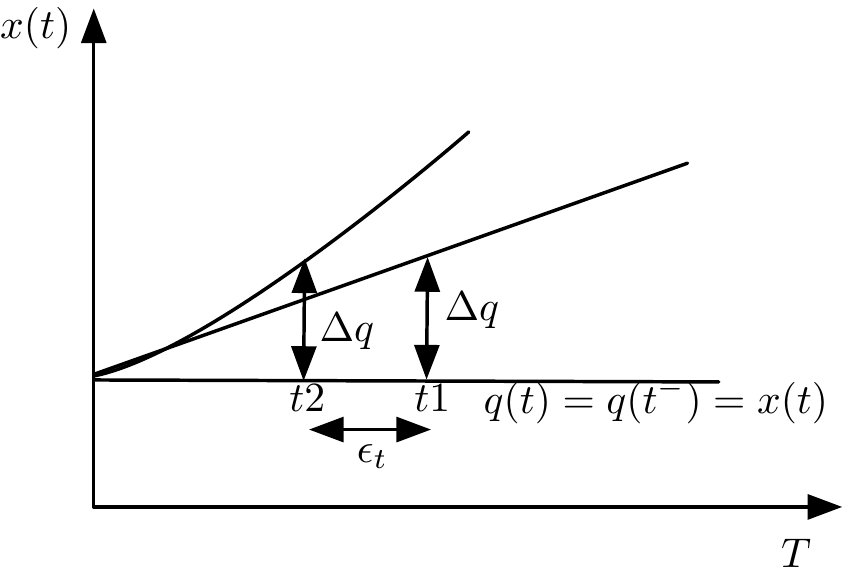} }
  \qquad
  \subfloat[The output trace for nonholonomic robot from the proposed
  technique. The trace \textit{always} converges to the level-crossing,
  but never crosses it like in other techniques.\label{fig:robottrace}]{
    \includegraphics[scale=0.5]{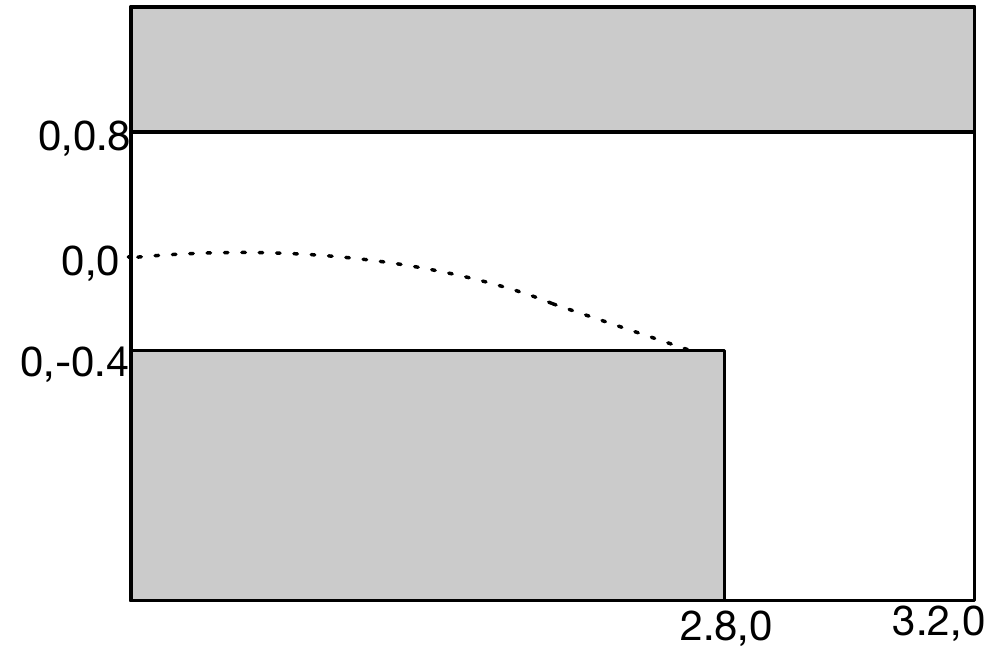}
  }
  \caption{Dynamic quantum selection and scheduling event generation}
  \label{fig:nextEvent}
\end{figure}

\begin{algorithm}[tb]
  \SetKwInOut{Input}{Input}\SetKwInOut{Output}{Output}
  \Input{Ode} 
  \Input{currentValues} 
  \Input{levelCrossings} 
  \tcp{The next time event $t$ to schedule \ac{QSHA}}
  \Output{$\mathbb{R}$} 
  \eIf{$0 \in \mathrm{levelCrossings}$}{
    \Return{$0$} \label{nE1}
  }
  {
    \ForEach{$l \in levelCrossings$}{ \label{lst:3}
      $toRet \leftarrow toRet \cup \{\mathrm{DQSS}(\mathrm{Ode},
      \mathrm{currentValues}, l)\}$\;\label{lst:5}
    }
  }
  \tcp{Return the very first integration step}
  \Return{$min(toRet)$\label{nE2}}
  \caption{The nextEvent algorithm}
  \label{alg:nextEventcode}
\end{algorithm}

\begin{algorithm}[t]
  \SetKwFunction{nextEvent}{\textrm{DQSS}}
  \nextEvent{$\textrm{Ode, currentVals, quanta}$}\label{DQSS1}\\
  \tcp{Compute $t1$ using QSS-1 Section~\ref{sec:acqss-1}}
  $t1 \leftarrow QSS1({Ode, currentVals, quanta})$\;
  \tcp{Compute $t2$ using $\Delta q = \mathrm{n\ degree\ polynomial\ q(t)} - q(t)^{-}$}
  $t2 \leftarrow MQSS2({Ode, currentVals, quanta})$\;
  \tcp{If $t1-t2 < \epsilon_{t}$ then return the value of $t1$}
  \tcp{\textbf{Dynamic\ selection\ of $\Delta q$}}
  \eIf{$abs(t1 - t2) < \epsilon_{t}$}{
    \Return{$t1$} \label{DQSS3}
  }
  {
    \Return{\nextEvent{$\textrm{Ode, currentVals, quanta/2}$}\label{DQSS2}}
  }
  \caption{The DQSS algorithm, computing the next time event $t$}
  \label{alg:DQSS}
\end{algorithm}

Algorithms~\ref{alg:nextEventcode} provides the pseudocode for computing
a time instant when the next integration step is scheduled. The
presented technique leverages \ac{QSS}-1 and a revised version of
\ac{QSS}-2 integration techniques. Figure~\ref{fig:nextEvent} visualises
the key idea of the proposed integration technique. Based on a given
level crossing value for a continuous variable $x$, we select an initial
guess of the level crossing as the $\Delta q=\mathbf{L_x} -\mathbf{x}$ where
$\mathbf{L_x}$ is the value of the level crossing. Usually
$\mathbf{L}_{x}$ is the value of the continuous variable $x$ that
satisfies one ore more of the outgoing edge guard condition from any
location of the \ac{HA}. Furthermore, $\mathbf{x}$ is the current value
of $x$. We then perform \ac{QSS}-1 integration to compute the time
instant (discrete event) $t1$ where $x(t) - q(t)=\Delta_q$, where $q(t)$ is
held constant using hysterisis. We also compute a time instant $t2$
where $x(t) - q(t)=\Delta_q$ where $x(t)$ is computed using a linear
approximation while $q(t)$ is held constant like before (this is a
variant of \ac{QSS}-2). If $t2-t1 > \epsilon_t$ then the selected quantum
$\Delta_q$ is too large as the error in the \ac{QSS}-1 approximation is too
large. We then halve the quantum and repeat the above steps until the
required error bound is achieved. Algorithm~\ref{alg:DQSS}, which
performs the above tasks recursively, is called by \texttt{nextEvent}
algorithm described below.

The \texttt{nextEvent} algorithm (Algorithm~\ref{alg:nextEventcode})
takes as input three arguments: \textcircled{1} The \ac{ODE} of the
continuous variable being used to compute the scheduling event,
\textcircled{2} the current values of all continuous variables in the
\ac{QSHA}, and \textcircled{3} a list of level crossings for this
continuous variable. For example, in location $\mathbf{T1}$, in
Figure~\ref{fig:robotha}, the variable $x$ has two possible values,
$3.2$ and $2.8$, that can satisfy guard $g$. Hence, the input
$\mathrm{levelCrossings}$ is the vector
$[3.2-\mathbf{x}, 2.8-\mathbf{x}]$, where $\mathbf{x}$ is the current
value of the continuous variable $x$. The algorithm returns a single
real number, which represents the future time event when the \ac{QSHA}
maybe executed.

Algorithm~\ref{alg:nextEventcode} first checks if any of the level
crossings has already happened (line~\ref{nE1}). If a level crossing has
already happened, the \ac{QSHA} is expected to be scheduled right this
instant and hence, the future integration step of 0 is returned.

If the level crossing has not yet occurred, then the \ac{DQSS}
(Algorithm~\ref{alg:DQSS}) is called to determine the next time event to
schedule the \ac{QSHA} for \textit{each} level in the
$\mathrm{levelCrossings}$ input (lines~\ref{lst:3}-\ref{lst:5}). Notice
that the last argument of Algorithm~\ref{alg:DQSS} is the quanta
(line~\ref{DQSS1}), which is set to the level crossing of the variable
(line~\ref{lst:5}, Algorithm~\ref{alg:nextEventcode}), when called from
Algorithm~\ref{alg:nextEventcode}. The \textit{minimum} from amongst all
the returned values from Algorithm~\ref{alg:DQSS} (line~\ref{nE2}) is
returned for scheduling the \ac{QSHA}.


\ignore{
The pseudocode of function \ac{DQSS} can be described using
Figure~\ref{fig:nextEventd}. We first find the time instant $t1$, where
the linear approximation of the state variable differs from the
quantized variable by the quantum $\Delta q$ exactly as in \ac{QSS}-1.
Next, we find the time instant $t2$, where the polynomial approximation
of the state variable differs from the quantized variable, held constant
to the current value of the state variable using hysteresis, by the
quantum $\Delta q$. If the difference between $t1$ and $t2$ is within
the user specified error bound $\epsilon_{t}$, then $t1$ is returned, so
that the \ac{QSHA} can be scheduled at $t1$ time units from now. If the
two time instants are too far apart, greater than user specified
$\epsilon_{t}$, then the quantum value is halved as seen on
line~\ref{DQSS2} and the \ac{DQSS} function is called with this smaller
quanta recursively. Note that the proposed \ac{DQSS} technique
dynamically (during simulation) selects the quanta and computes the
resultant discrete time event $t$, unlike the state-of-the-art \ac{QSS}
techniques~\cite{kofman2004discrete,kofman2006third}, which use a static
quantum.}




\subsection{Simulation of \ac{QSHA}}

\begin{algorithm}[tb]
  \SetKw{Break}{break}
  \SetKwInOut{Input}{Input}\SetKwInOut{Output}{Output}
  \Input{\ac{QSHA}
    $\mathcal{H}_q = \langle L, X,
    X_q, Init, f, f_{q}, h, Inv, E, G, R \rangle$}
  \Input{Simulate until time $U$}
  \Output{Trace $\Gamma$}
  \tcp{Current simulation time}
  \KwData{$T_{n} \leftarrow 0$}
  \tcp{Previous value of $T_{n}$}
  \KwData{$T_{pre} \leftarrow 0$}
  \tcc{The next event time}
  \KwData{$t \leftarrow 0$} 
  \tcp{Previous value of continuous variables}
  \KwData{$X_{pre} \leftarrow \{0 | x \in X\}$}
  \tcp{Current value of continuous variables}
  \KwData{$X_{now} \leftarrow$ \{$0 | x \in X$\}}
  \tcc{Location enabled for execution}
  \KwData{$enable \leftarrow l_{0}$}
  \tcc{Initialise the pre-form variables}
  $X_{pre} \leftarrow \{\mathbf{x} | 
    (enable, \mathbf{x}) \in Init$\}\label{o1}

\ignore{
  \lForEach{$x \in X$}{
    $X_{pre}[x] \leftarrow \{\mathbb{R} | \mathbb{R}
    \in Init(enable, x)$\}\label{o1}}
  }
  \tcc{While simulation time has not exceeded time $U$}
  \While{$T_{n} \leq U$ \label{o3}}{
    \tcp{For each enabled location}
    TOP: \For{$l \in L \wedge enable == l$ \label{top}}{
      \tcc{All outgoing edges of location $l$}
      \ForEach{$e \in E \wedge e(0) == l$ \label{o4}}{
        \tcc{pre variables satisfy guard --- inter-location transition}
        $\hat{X_{pre}} = \{\tilde{X_{pre}} \in 2^{X_{pre}}
        |  \tilde{X_{pre}} \subseteq G(e)\}$\label{o5}\;
        \If{$\hat{X_{pre}} \neq \emptyset $ \label{o52}}{
          $X_{now} \leftarrow \{\mathbf{x} | \mathbf{x} \in R(e, {X_{pre}})$\}\label{o6}\\
          \ignore{
          $\hat{X_{now}} \leftarrow \{\mathbb{R}^{|X|} |
          \mathbb{R}^{|X|} \subseteq R(e, X_{pre})\}$\;
          \lForEach{$x \in X_{now}$}{
            $X_{now}[x] \leftarrow \hat{X_{now}}[x]$}
          }
          \tcc{Update the next enabled location}
          $enable \leftarrow e(1)$\label{o7}\;
          \tcc{location switch is instantaneous and hence $t$ is set to
          $0$}
          $t \leftarrow 0$\label{o8}\;
          \tcc{Jump out of loop}
          \Break TOP \label{ob}\label{o9}\;
        }
        \label{o4e}}
      {
        \tcc{Evolve \acp{ODE} in location $l$ using Euler integration---
          intra-location transition}
        \tcp{For the very first time, $T_{n} = T_{pre} = 0$}
        $\delta \leftarrow T_{n} - T_{pre}$ \label{o10}\;
        $X_{now} \leftarrow X_{pre}+ f(l, X_{pre}) \times \delta$\label{o11}\;
        \tcc{Compute next discrete event $t$}
        $t \leftarrow \mathrm{compute\_next\_event}(\mathcal{H}_q, l, X_{now})$\label{o13n}\;
      }
    }\label{o13}
    \tcc{Update the pre variables}
    $X_{pre} \leftarrow X_{now}$\label{o20}\;
    \tcc{Concatenate to output trace} 
    $\Gamma \leftarrow \Gamma \cup \{(T_{n}, enable, X_{pre})\}$\label{o21}\;
    \tcc{Save current time $T_{n}$}
    $T_{pre} \leftarrow T_{n}$\label{o22}\;
    \tcc{Increment the simulation time by $t$}
    $T_{n} \leftarrow T_{n} + t$\label{o23}\;
  }
  \Return{$\Gamma$}
  \caption{Top-level Discrete Event Simulation engine for \ac{QSHA}}
  \label{alg:discr-event-simul}
\end{algorithm}


Algorithm~\ref{alg:discr-event-simul} describes the top-level discrete
event simulation engine for simulating the \ac{QSHA}. The algorithm
takes as input the \ac{QSHA} that needs to be simulated
($\mathcal{H}_q$) and the total time for simulation ($U$) and produces
the output trace $\Gamma$. The algorithm starts be declaring two counters
($T_{n}$ and $T_{pre}$) for holding the current instant of execution and
previous instant of execution of $\mathcal{H}_q$ respectively. Variable
$t$ represents the duration of the discrete event simulation step.
Variables $X_{now}$ and $X_{pre}$ represent the values of the continuous
variables of the \ac{QSHA} at time $T_{n}$ and $T_{pre}$
respectively. 
Variable $enable$ holds the current enabled location of $\mathcal{H}_q$,
which is initialized to the the initial location $l_{0}$ at the start of
simulation i.e. $T_{n} = 0$.

\ignore{Variables $X_{now}$ and $X_{pre}$
are dictionaries, i.e., key-value pairs, where the keys are the
continuous variable names. A specific variable value may be obtained
(or set) from the dictionary using $X_{now}[x]$, where $x$ is a
continuous variable. The domains of these variables are represented
as $\mathbf{X_{now}}$ and $\mathbf{X_{pre}}$ respectively. Variable
$enable$ holds the current enabled location from the \ac{QSHA}, which
is initialized to the the initial location $l_{0}$ at the start of
simulation i.e. $T_{n} = 0$.}

Algorithm~\ref{alg:discr-event-simul} starts by initializing all the
variables in dictionary $X_{pre}$ with the initial values specified by
$Init$ of the \ac{QSHA} (line~\ref{o1}). For the nonholonomic robot
running example (Figure~\ref{fig:robotha}), line~\ref{o1} in the
algorithm would initialise $x = y = \theta = 0$ and $\phi = 1$ in the set
$X_{pre}$. The output trace $\Gamma$ is always concatenated with the values
of the variables from $X_{pre}$ along with the current time $T_{n}$ and
the enabled location (line~\ref{o21}).

The main simulation engine is a loop that iterates, executing the
\ac{QSHA} in discrete time steps until the current simulation time is
less than or equal to the total user specified simulation time
(line~\ref{o3}). The execution of the \ac{QSHA} is carried out as
follows:

\begin{enumerate}
\item From any enabled location, first the guards on all outgoing edges
  are checked. If any guard is \texttt{True}, then the outgoing edge is
  taken instantaneously in zero simulation time. This behaviour
  corresponds to the inter-location transition described in
  Section~\ref{sec:syntax-semantics} and implemented from
  lines~\ref{o4}-\ref{o4e} in Algorithm~\ref{alg:discr-event-simul}. For
  the running example of the nonholonomic robot in
  Figure~\ref{fig:robotha}, location \textbf{T1} is selected by
  line~\ref{top} as being enabled initially. Next, the loop body
  (lines~\ref{o5} and~\ref{o52}) checks if the values in $X_{pre}$
  satisfy the outgoing edge guard $g$ from Figure~\ref{fig:robotha}. If
  the guard condition is satisfied, then the variables $X_{now}$ are
  updated according to the reset relation on the outgoing edge
  (line~\ref{o6}). Finally, the next destination location is enabled
  (line~\ref{o7}), the next event variable $t$ is set to zero
  (line~\ref{o8}) and the algorithm iterates after updating the pre set
  of variables, the current simulation timer, and the output trace
  (lines~\ref{o9} and~\ref{o20}-\ref{o23}).

\item If none of the outgoing edge guards hold then, \textcircled{a} the
  \acp{ODE} in the enabled location of a \ac{QSHA} evolve
  (lines~\ref{o10} and~\ref{o11}) and \textcircled{b} the
  \textit{future} discrete time event $t$, when the \ac{QSHA} will be
  executed next is computed (line~\ref{o13n}) --- this behaviour
  corresponds to the intra-location transition described in
  Section~\ref{sec:syntax-semantics}. During evolution of the \ac{ODE}
  and computation of the future discrete event $t$ the following actions
  are performed:

  \begin{enumerate}
  \item The values of the continuous variables are updated using the
    standard forward Euler numerical integration technique. This
    integration uses the values from $X_{pre}$ and the updates are
    performed on the variables of $X_{now}$, i.e., the right hand-side
    of all updates work on the variables from the $X_{pre}$, while the
    left hand side updates are always performed on the $X_{now}$. In
    case of the running example, the nonholonomic robot, this step would
    evolve the variables $x$, $y$, $\theta$, and $\phi$ in location
    $\mathbf{T1}$ or $\mathbf{T2}$. The very first iteration of the
    algorithm, which evolves the \acp{ODE} has a time step ($\delta$) of zero
    (line~\ref{o10}), because for the very first time
    $T_{n} = T_{pre} = 0$. Thus, the variables will not change their
    values at the start, until the duration of non-zero integration step
    is decided (see below).

  \item The calculation of the future discrete simulation event $t$ is
    carried out by function $\mathrm{compute\_next\_event}$ described in
    Algorithm~\ref{alg:callnE}. This function takes as input the
    \ac{QSHA} ($\mathcal{H}_q$), the current enabled location $l$, and
    $X_{now}$, which contains the current values of the continuous
    variables. This next discrete time event $t$ returned by this
    function is used to schedule the \ac{QSHA} in the next iteration of
    the algorithm. The $\mathrm{compute\_next\_event}$ function is
    described in the next section.

  \end{enumerate}
\end{enumerate}

\subsection{Computing the next discrete simulation (integration) step}
\label{sec:computeNE}

\begin{algorithm}
  \SetKwInOut{Input}{Input}\SetKwInOut{Output}{Output}
  \Input{\ac{QSHA}
    $\mathcal{H}_{q} = \langle L, X,
    X_q, Init, f, f_{q}, h, Inv, E, G, R \rangle$}
  \tcp{The current enabled location}
  \Input{$enabled \in L$}
  \tcp{The current value of continuous variables}
  \Input{$X_{now}$}
  \tcp{The next discrete simulation/integration event $t$}
  \Output{$t: \mathbb{R}^{\geq 0}$}

  \tcp{Initialize set $\tilde{ts}$}
  $\tilde{ts} \leftarrow \emptyset$\label{o12}\;
  \ForEach{$e \in E \wedge e(0) == enabled$}{
    \tcc{$LC$ is a $M \times |X|$ matrix, where each row indicates the
      value of continuous variables that satisfy the outgoing edge $e$
      guard $G (e)$}
    $LC \leftarrow [ \mathbb{R}^{|X|} |
      \forall R^{|X|} \in 2^{\mathbb{R}^{|X|}}
      \wedge R^{|X|} \subseteq  G(e) ]$\label{lc1}\;
    \ForEach{$x \in X$ \label{ne1}}{
      $ode_{x} \leftarrow f_{q}(l, X_{now}[x])$\;
      \tcp{Index through each column in $LC$}
      $LC_{x} \leftarrow \{\mathbb{R}-X_{now}[x] | \mathbb{R} \in LC^{T}[x] \}$\;
      \tcc{If guards depends upon $x$}
      \If{$LC_{x} \neq \infty$}{
        $\tilde{ts} \leftarrow \tilde{ts} \cup \{\mathrm{nextEvent}(ode_{x}, X_{now},
        LC_{x})\}$\;
      }
    }
    $t \leftarrow min(\tilde{ts})$\label{o14}\;
    \Return{t}
  }
  \caption{Function $\mathrm{compute\_next\_event}$ that computes the next
    discrete simulation/integration step event}
  \label{alg:callnE}  
\end{algorithm}

Algorithm~\ref{alg:callnE} describes the function that computes the next
discrete simulation/integration step. This function takes as input the
\ac{QSHA} $\mathcal{H}_{q}$, the currently enabled location $enabled$,
and the current valuation of the continuous variables $X_{now}$. The
function returns a real value $t$, which indicates the next integration
and discrete event simulation step.

Algorithm~\ref{alg:callnE} first initializes a set $\tilde{ts}$
(line~\ref{o12}), which will hold the future event for each variable
used in the guard $g$ of the outgoing edge of the currently enabled
location. In case of the nonholonomic robot, two variables $x$ and $y$
are used in the guard $g$ from location $\mathbf{T1}$. Hence, the set
$\tilde{ts}$ will hold two future time events. Next, the algorithm
computes the matrix $LC$, which satisfies the guard on each outgoing
edge of the enabled location (line~\ref{lc1}). In case of the
nonholonomic robot, with $\mathbf{T1}$ as the enabled location, with a
single outgoing edge, the matrix $LC$ is given by
Equation~(\ref{eq:lc1}). The columns in Equation~(\ref{eq:lc1}),
represent $x$, $y$, $\theta$, and $\phi$, respectively. Notice that
$\mathbb{R}$ denotes that the guard is satisfied for any value of that
variable, i.e., the guard is \textit{not} affected by that variable.

\begin{equation}
  LC = 
  \begin{bmatrix}
    x & y & \theta & \phi \\
    \hline
    3.2 & 0.8 & \mathbb{R} & \mathbb{R}\\
    2.8 & -0.4 & \mathbb{R} & \mathbb{R}\\
  \end{bmatrix}
  \label{eq:lc1}
\end{equation}

Finally, for each continuous variable, if the guard on the outgoing edge
depends upon that variable, the $\mathrm{nextEvent}$ function calculates
the future discrete event to schedule the \ac{QSHA}
(lines~\ref{ne1}-\ref{o13}). In case of the nonholonomic robot, the set
$\tilde{ts}$ contains two possible real values when the \ac{QSHA} can be
scheduled, because the guard only depends upon variables $x$ and $y$.
The last argument ($LC_{x}$) to function $\mathrm{nextEvent}$, is a
vector of real values indicating the \textit{difference} between the
values of variables that satisfy the guard and the current value of the
continuous variables. For example, when calling $\mathrm{nextEvent}$ for
variable $x$ $LC_{x} = [3.2-\mathbf{x}, 2.8-\mathbf{x}]$, where
$\mathbf{x}$ indicates the current value of variable $x$. The algorithm
returns the minimum from amongst all the values in $\tilde{ts}$ for
scheduling the \ac{QSHA}.

\subsection{Properties of the proposed dynamic quantum based
  simulation}
\label{sec:prop-prop-dynam}

There are a number of useful properties that the proposed \ac{DQSS}
simulation technique enforces.

\paragraph{Simulation progress}

As described in Section~\ref{sec:comp-next-sched}, we find two time
instants $t1$ and $t2$, for \ac{QSS}-1 and revised \ac{QSS}-2, when
selecting the quantum $\Delta q$ dynamically. These time instants are roots
of polynomials of some degree $n$. In case of the running example, for
instance, solving \ac{QSS}-1 results in a linear equation (polynomial of
degree 1, c.f. Section~\ref{sec:acqss-1}), while \ac{QSS}-2 results in a
polynomial with a transcendental function (c.f. Equation~(\ref{eq:1})
c.f. Section~\ref{sec:acqss-2}). This transcendental function needs to
be expanded to a Taylor series, with finite terms, resulting in a
polynomial of degree $n$. The roots of these polynomials are the
discrete time events when the \ac{QSHA} can be scheduled. Our technique
always results in at least one real positive root for the $n$ degree
polynomial (Equation~(\ref{eq:4})), which we prove in
Theorem~\ref{sec:theorem-real-postitive}\footnote{This property is not
  guaranteed for standard \ac{QSS}-2 and higher degree \ac{QSS}
  techniques}. This property guarantees that simulation always makes
progress.

\begin{theorem}
  \label{sec:theorem-real-postitive}
  Let
  $f(x) = a_{n}x^{n} + a_{n-1}x^{n-1} + \ldots + ax \pm \Delta q = 0$,
  be a polynomial of degree $n$ with real coefficients. Then $f(x)$ has
  \textit{at least} one \textit{real positive} root.
\end{theorem}
\begin{proof}
  Let $v = n - l$, $\forall l \in [1, n]$ be the number of sign changes
  of the sequence $(a_{n},\ldots,a)$. Then by Descartes's rule of
  signs~\cite{curtiss1918recent} we have $p = v - 2m$, where $p$ is the
  number of real positive roots for some non-negative integer $m$, i.e.,
  $m \in \mathbb{Z}^{\geq 0}$. From fundamental theorem of algebra we
  have $p \geq 0$. Thus, if a polynomial has no real positive roots,
  $p = 0$. This can happen only when $m = \frac{v}{2} = \frac{n-l}{2}$,
  for polynomial with coefficients $(a_{n},\ldots,a)$.

  Case 1: $n-l$ is an even number. We can always make it an odd number
  by adding $\Delta q$, which has the opposite sign to that of
  coefficient $a$. Hence, for polynomial $f(x)$, we get
  $m = \frac{(n-l+1)}{2}$ if there are no real positive roots. But,
  $n-l+1$ is odd, because $n-l$ is even. Thus,
  $m \notin \mathbb{Z}^{\geq 0}$, which violates Descartes's rule. The
  value that $m$ can take for minimum $p$, such that $p \geq 0$ and
  $m \in \mathbb{Z}^{\geq 0}$ is one less than $\frac{(n - l + 1)}{2}$,
  which is $\frac{n - l}{2}$. Hence,
  $p = {n-l+1} - (2(\frac{n-l}{2})) = 1$. Hence, there is at least one
  real positive root of the polynomial $f(x)$.

  Case 2: $n-l$ is an odd number. We can enforce that $n-l$ remains odd,
  by adding $\Delta q$ with the same sign as that of coefficient $a$.
  Hence, for polynomial $f(x)$ to have zero real positive roots, we have
  $m = \frac{n-l}{2}$. Here again, since $n-l$ is odd,
  $m \notin \mathbb{Z}^{\geq0}$. The value that $m$ can take for minimum
  $p$, such that $p \geq 0$ and $m \in \mathbb{Z}^{\geq 0}$ is
  $\frac{(n - l - 1)}{2}$. Giving us
  $p = {n-l} - (2(\frac{n-l-1}{2})) = 1$. Hence once again there is at
  least one real positive root of the polynomial $f(x)$.
  
\end{proof}

\paragraph{The simulation always detects the first level crossing}

Algorithm~\ref{alg:discr-event-simul} line~\ref{o10} uses forward Euler
to evolve the \acp{ODE} using the time-step returned by \ac{QSS}-1
(Algorithm~\ref{alg:DQSS} line~\ref{DQSS3}). First we show that this is
correct, i.e., from any given point in time $T_{n}$ if we use forward
Euler to compute the integral ($x(T_{n} + t)$) of any function $x(t)$,
where time-step $t$ is computed using \ac{QSS}-1 with a quantum
$\Delta q$, then $|x (T_{n}+t)-x (T_{n})| = \Delta q$.

\begin{lemma}
  For some function $x(t)$, let $\dot{{x}} = f(x(t),t)$ be its slope.
  Then, given $x (T_{n} + t) = \dot{x}\times t + x (T_{n})$ by forward Euler,
  where $t$ is obtained by \ac{QSS}-1 for function $x(t)$ using a
  quantum of $\Delta q$, $|x(T_{n}+t)-x (T_{n})| = \Delta q$.
  \label{lem:1}
\end{lemma}
\begin{proof}
  Simplifying our goal, we have:
  \begin{align}
    x (T_{n}+t) = \pm \Delta q + x (T_{n})\\
    \therefore f(x(T_{n}), T_{n}) \times t + x (T_{n}) = \pm \Delta q + x (T_{n})\\
    \therefore f(x(T_{n}), T_{n}) = \pm \Delta q/t\label{eq:ts}
  \end{align}

  By definition of \ac{QSS} we have:
  \begin{align}
    \dot{x} = f(q(t), t)
  \end{align}

  Integrating $x(t)$ using forward Euler from $T_{n}$ to $t$ with the
  \ac{QSS} approximation we have:
  \begin{align}
    x(T_{n}+t) = f(q(t), t)\times t + x (T_{n})
  \end{align}
  By hysteresis in \ac{QSS}-1 we have:
  \begin{align}
    q(t) = q(t^{-}) = x(T_{n})\\
    \therefore x (T_{n} + t) = f (x(T_{n}),T_{n}) \times t + x (T_{n})
  \end{align}
  By \ac{QSS}-1, we have
  \begin{align}
    x (T_{n}+t)-q(t) = \pm \Delta q\\
    \therefore \mathrm{by\ hysteresis}\ x (T_{n}+t)-x(T_{n}) = \pm \Delta q\\
    \therefore f (x(T_{n}, t)\times t + x(T_{n}) - x (T_{n}) = \pm \Delta q\\
    \therefore t = \pm \Delta q/f (x(T_{n}), t)\label{eq:t}
  \end{align}

  Substituting Equation~\ref{eq:t} in Equation~\ref{eq:ts} we have
  \begin{align}
    f(x(T_{n}), T_{n}) = \pm \Delta q/ (\pm \Delta q/f (x(T_{n}, T_{n}))\\
    \therefore f(x(T_{n}), T_{n}) = f (x (T_{n}), T_{n})
  \end{align}
\end{proof}

Theorems~\ref{thm:2} and~\ref{thm:3} together show that
Algorithm~\ref{alg:discr-event-simul} always converges to the first
level crossing from any given location in the \ac{QSHA}.

\begin{theorem}
  Given a \ac{QSHA}
  $\mathcal{H}_{q} = \langle L, X, X_{q}, Init, f, f_{q}, h, Inv, E, G, R\rangle$,
  $\exists t \in \mathbb{R}$, such that
  ${x}(t) \in G(e), \forall e \in E, \forall x \in X_{q}$ and
  $\forall T_{n} \in [0, t)$,
  ${x}(T_{n}) \notin G(e), \forall e \in E, \forall x \in X_{q}$. Then, in any given location
  $l \in L$,
  $\int^{T_n+\delta}_{T_n}\dot{x}(\tau)d\tau \in G(e), \forall x \in X$, i.e., our simulation
  technique can take an integration step $\delta$, such that the level
  crossing is satisfied and $T_n + \delta = t$.
  \label{thm:2}
\end{theorem}

\begin{proof}
  Without loss of generality we consider each continuous variable
  $x \in X_{q}$ separately. Note that our integration technique always
  returns the next time event \texttt{t1}, which corresponds to the
  \ac{QSS}-1 integration step (Algorithm~\ref{alg:DQSS},
  line~\ref{DQSS3}). Hence, we apply the definition of \ac{QSS}-1 (c.f.
  Section~\ref{sec:acqss-1}) to prove this theorem. From \ac{QSS}-1, we
  have:
  \begin{align}
    |x(T_n+\delta) - q(T_n)| = \Delta q \\
    \therefore |x(T_n + \delta) - x(T_n)| = \Delta q, \mathrm{by\ hysteresis}\label{eq:2}\\
    \therefore x(T_n + \delta) - x(T_n) = \pm \Delta q
  \end{align}
  From Algorithms~\ref{alg:nextEventcode} and~\ref{alg:DQSS}, our
  simulation technique selects a \mbox{$\Delta q \leq x(t) - x(T_n)$}.

  Case 1: $\Delta q = x(t) - x(T_n)$. By forward Euler $x(T_n+\delta) =
  \delta\times \dot{x}(T_n) + x(T_n)$. Applying Equation~(\ref{eq:2}), we
  have
  \begin{align}
    \delta\times\dot{x}(T_n) + x(T_n) - x (T_n) = \pm \Delta q \\
    \therefore \delta = \pm \frac{\Delta q}{\dot{x}(T_n)}\label{eq:3}
  \end{align}
  We choose a positive or negative $\Delta q$, such that $\delta$ is
  always real positive (from Theorem~\ref{sec:theorem-real-postitive}).
  Hence, in this case, assuming $\dot{x}(T_{n})$ is positive, without
  loss of generality, and choosing a positive $\Delta q$, and
  substituting $\delta$ back in forward Euler, we have:
  \begin{align}
    x (T_n + \delta) = x (T_n) + \frac{\Delta q}{\dot{x} (T_n)} \times \dot{x} (T_n)\\
    \therefore x (T_n + \delta) = x (T_n) + \Delta q\\ 
    \therefore x (T_n + \delta) = x (T_n) + x (t) - x (T_n)\\
    \therefore x (T_n + \delta) = x (t)
  \end{align}

  Hence, $x (T_n + \delta) \in G (e)$ and by injective property of
  functions $T_n + \delta = t$.

  Case 2: $\Delta q < x (t) - x (T_n)$. From Equation~(\ref{eq:3}), we
  have some $\delta' < \delta$, where $\delta$ is given by Case 1.
  Hence, $T_n + \delta' < T_n + \delta$. Binding $T_{n}$ to
  $T_{n} + \delta'$, we need to prove that
  $\int^{T_n + \delta'}_{T_n}\dot{x} (\tau)d\tau \in G (e)$, which can
  be proven recursively from Case 1 and Case 2.
  
\end{proof}

From Equation~(\ref{eq:3}) it is clear that the discrete integration
(simulation) step size $\delta$ is directly proportional to the selected
quantum $\Delta q$. As the current values ($x (T_{n})$) of the continuous
variables get closer to the level crossings, i.e., values that satisfy
the outgoing edge guards ($x(t)$), $\Delta q$ becomes smaller, because
$\Delta q \leq x(t)-x(T_{n})$. Consequently, the integration step size also
becomes smaller as we approach the level crossings, which is as shown in
Figure~\ref{fig:timeset}.

\begin{theorem}
  Given a well-formed \ac{HA} and consequently a well-formed \ac{QSHA}
  $\mathcal{H}_{q} = \langle L, X, X_{q}, Init, f, f_{q}, h, Inv, E, G, R\rangle$
  Algorithm~\ref{alg:discr-event-simul} always converges to a time event
  $t$, such that $\forall x \in X, x(t)$ satisfies $G(e)$ and
  $\forall t' < t$, $x(t)$ does not satisfy $G(e)$, $\forall e \in E$.
  \label{thm:3}
\end{theorem}
\begin{proof}
  We will consider the case where the guard for any edge $e \in E$ is
  only dependent upon two variables $x, y \in X$, and the proof can be
  generalized to any number of variables.

  Case 1: We consider the case where the guard is a disjunction of two
  variables $x$ and $y$ of the form:
  $l_{x}\ \tilde{\bowtie}\ x\ \tilde{\bowtie}\ u_{x} \vee l_{y}\
  \tilde{\bowtie}\ y\ \tilde{\bowtie}\ u_{y}$, where
  $\tilde{\bowtie} \in \{<, \leq\}$. We need to show that
  Algorithm~\ref{alg:discr-event-simul} always finds the time instant
  $t$ such that $x(t) \in [l_{x}, u_{x}] \vee y(t) \in [l_{y}, u_{y}]$
  and $\forall t' \in [0, t)$ the guard is not satisfied.

  Case a: There exists some $t^{l}_{x} < t^{l}_{y}$ such that
  $x(t^{l}_{x})-l_{x} = 0$ and
  $\forall t' \in [0, t_{x}), x (t)-l_{x} \neq 0$, from Theorem~\ref{thm:2}, where
  $t^{l}_{y}$ is the first time instant where $y(t^{l}_{y})-l_{y} = 0$
  again from Theorem~\ref{thm:2}. Hence,
  $x(t^{l}_{x}) \in [l_{x}, u_{x}]$. By definition of disjunction
  $x(t^{l}_{x}) \in [l_{x}, u_{x}] \vee y(t^{l}_{x}) \in [l_{x}, u_{x}]$.
  Algorithm~\ref{alg:discr-event-simul} always selects the minimum from
  amongst all the times for each variable (line~\ref{o14}). Hence, is
  satisfies the requirement $t^{l}_{x} < t^{l}_{y}$.

  Algorithm~\ref{alg:discr-event-simul} always finds the time
  $t = t^{l}_{x} = t^{l}_{y}$ and
  $t = t^{l}_{y} s.t., t^{l}_{y} < t^{l}_{x}$ when the outgoing edge
  guard holds follows the same reasoning as case a.

  Case 2: We consider the case where the guard is a conjunction of two
  variables $x$ and $y$ of the form:
  $l_{x}\ \tilde{\bowtie}\ x\ \tilde{\bowtie}\ u_{x} \wedge l_{y}\
  \tilde{\bowtie}\ y\ \tilde{\bowtie}\ u_{y}$, where
  $\tilde{\bowtie} \in \{<, \leq\}$. We need to show that
  Algorithm~\ref{alg:discr-event-simul} always finds the time instant
  $t$ such that $x(t) \in [l_{x}, u_{x}] \wedge y(t) \in [l_{y}, u_{y}]$
  and $\forall t' \in [0, t)$ the guard is not satisfied.

  Case a: There exists $t^{l}_{x} < t^{l}_{y}$ such that
  $x (t^{l}_{x})-l_{x} = 0$ and
  $\forall t' \in [0, t_{x}), x(t)-l_{x} \neq 0$, from Theorem~\ref{thm:2}, where
  $t^{l}_{y}$ is the first time instant where $y(t^{l}_{y})-l_{y} = 0$
  again from Theorem~\ref{thm:2}. Let $t_{x}^{u}$, be the first time
  instant where $x (t_{x}^{u})-u_{x}=0$ from Theorem~\ref{thm:2}. If
  $t^{l}_{x} < t^{l}_{y} \leq t_{x}^{u}$, then the first time instant
  $t$ where the conjunctive guard is satisfied is
  $[t^{l}_{x}, t_{x}^{u}] \cap \{t^{l}_{y}\} = t^{l}_{y}$. Since
  Algorithm~\ref{alg:discr-event-simul} selects the minimum time from
  amongst level crossing times for all variables (line~\ref{o14}), it
  always finds time $t = t^{l}_{y}$. Case $t^{l}_{y} > t_{x}^{u}$ cannot
  happen, because then the \ac{QSHA} is not well-formed.

  Algorithm~\ref{alg:discr-event-simul} finds the time instant
  $t = t^{l}_{x} = t^{l}_{y}$ and
  $t = t^{l}_{y}, s.t., t^{l}_{y} < t^{l}_{x}$ follows the same
  reasoning as case a.
\end{proof}

Figure~\ref{fig:robottrace} gives the resultant trace of the
nonholonomic robot. Notice that the \ac{QSHA} always converges to the
level crossing, i.e., the obstacle, but never overshoots it.

%% file: exp.tex
\section{Experimental results}
\label{sec:experimental-results}

This section describes the experimental set-up and results that we use
to quantitatively validate the proposed discrete event simulation
technique for \ac{QSHA}.

\begin{table}[tb]
  \caption{Benchmark description}
  \label{tab:benchmarks}
  \centering
  \begin{tabular}{|c|c|c|p{30pt}|p{40pt}|p{30pt}|}
    \hline
    Benchmarks& \#L & \#\acp{ODE}& \ac{ODE} types& Level-crossing interval & Simulation Time (sec). \\
    \hline
    TH& 2 & 2& Linear & Open & 0.5 \\
    \hline
    WLM& 4 & 8 & Linear & Open & 30 \\
    \hline
    Robot&  2 & 8 & Non-Linear & Open, logically combined & 0.07 \\
    \hline
    AFb&  4 & 20 & Non-Linear & Open & 1.6 \\
    \hline
    Reactor&  4 & 9 & Linear & Closed, logically combined & 30 \\
    \hline
  \end{tabular}
\end{table}

\subsection{Benchmark description}
\label{sec:benchmarkdesc}

We have selected a number of different published benchmarks from
different domains and complexity to validate the proposed technique. A
quick overview of the selected benchmarks is provided in
Table~\ref{tab:benchmarks}. These \acp{HA} are converted into \acp{QSHA}
before simulation. We give a more detailed overview of the benchmarks
here in:

\begin{itemize}
\item \ac{TH}: The \ac{TH}~\cite{ye2008modelling} benchmark is a \ac{HA}
  with two locations (indicated by \#L in Table~\ref{tab:benchmarks})
  with one \ac{ODE} in each location. \ac{TH} benchmark describes
  maintaining the temperature in a room between an upper and lower
  bound. The \acp{ODE} are linear, of the form $\dot{x} = f(x(t), t)$.
  The guards on \ac{HA}' edges (level crossings) are open intervals,
  e.g., $x \geq \theta$ or $x \leq \theta$. After 0.5 seconds of
  simulation time the \ac{HA} reaches a steady state with a repeating
  trace.

\item \ac{WLM}: The \ac{WLM}~\cite{alur1995algorithmic} example is a
  \ac{HA} that maintains the water in tanks within some level. This
  \ac{HA} contains four locations (indicated by \#L in
  Table~\ref{tab:benchmarks}) with two \acp{ODE} in each location. The
  \acp{ODE} have a constant slope. The edge guards (level crossings) are
  open intervals like in \ac{TH}. After 30 seconds of simulation time,
  the \ac{HA} reaches a steady state with a repeating trace.

\item \ac{Robot}: The \ac{Robot}~\cite{de1998feedback} example is the
  running example from the paper. As seen from Figure~\ref{fig:robotha},
  there are two locations with 8 \acp{ODE} altogether. The \acp{ODE} in
  this example are non-linear. In fact, the slopes are transcendental
  (trigonometric) functions, which are \textit{approximated} with a
  Taylor polynomial. The guards on \ac{HA}' edges are open intervals,
  but dependent upon more than one variable. For example, the guard $g$
  in Figure~\ref{fig:robotha} depends upon two variables $x$ and $y$. We
  detect a collision with the objects within 0.07 simulation seconds
  when $v1 = 30\ \mathrm{m/s}$ and $v2 = -10\ \mathrm{m/s}$ in
  Figure~\ref{fig:robotha}.

\item \ac{AFb}: The \ac{AFb}~\cite{grosu2011cardiac} \ac{HA} simulates
  fibrillation (abnormal heart rhythm) in the human atria. There are a
  total of four locations with five \acp{ODE} in each location. The
  \acp{ODE} are coupled and complex of the form:
  \[
      \begin{aligned}
        \dot{u} = e + (u-\theta_v)(u_u-u ) v g_{fi} +
        wsg_{si}-g_{so}(u) \\
        \dot{s} =
        \displaystyle\frac{g_{s2}}{(1+\exp(-2k(u-us)))} - g_{s2}s \\ \
        \dot{v} = -g_v^+\cdot v \\
        \ \dot{w} = -g_w^+\cdot w
      \end{aligned}
  \]
  \noindent In the above equations $u$ indicates the transmembrane
  potential of the heart, $v$ is the fast channel gate, while $w$ and
  $s$ are slow channel gates for sodium, potassium, and calcium ionic
  flow. The rest are constant parameters. Once again the exponential
  function is \textit{approximated} with a Taylor polynomial. \ac{AFb}
  reaches a Zeno state after 1.6 seconds of simulation time.
  
\item \ac{Reactor}: \ac{Reactor}~\cite{jaffe1991software} \ac{HA}
  simulates cooling and control of a nuclear plant. It contains four
  locations, three of which have three \acp{ODE} each and the fourth
  location is a shut-down location. An interesting aspect of this
  benchmark is that the edge guards have closed interval of the form
  $x = y \wedge \tau = \theta$\footnote{Equality can be expressed using
    less than or equal to operators from Definition~\ref{def:guards}},
  i.e., there are equality constraints combined together with
  conjunction. Equality on level crossings in known to be difficult to
  deal with in most hybrid simulation frameworks. Furthermore this
  \ac{HA} is also \textit{non-deterministic}. We purposely chose this
  example to see how the other discrete event simulation tools handle
  both equality on edge guards and non-determinism. The \ac{Reactor} in
  this benchmark shuts-down after 30 seconds of simulation time.
\end{itemize}

\subsection{Tool set-up}
\label{sec:toolsetup}

We have compared the proposed discrete event simulation framework,
available from~\cite{eha}, with Ptolemy~\cite{ptolemaeus2014system},
which is a robust discrete event simulation framework for different
design paradigms, including \ac{HA}. Ptolemy allows simulating \acp{HA}
with
\ac{QSS}~\cite{LeeNiknamiNouiduiWetter15_ModelingSimulatingCyberPhysicalSystemsUsingCyPhySim}
and Runge-Kutta techniques. We also used
Modelica~\cite{tiller2012introduction} programming language and the
OpenModelica~\cite{fritzson2006openmodelica} simulation framework for
comparison purposes.

\begin{table}[tb]
  \caption{Tool set-up}
  \label{tab:techs}
  \centering
  \begin{tabular}{|c|c|c|c|c|}
    \hline
    Technique & $\Delta q$& $\epsilon_{v}$ & $\epsilon_{t}$& maximum $\Delta t$ \\
    \hline
    \ac{QSS}-1 & $10^{-3}$ & $10^{-6}$ & N/A & N/A \\
    \hline
    \acs{RK}-45 & N/A & $10^{-6}$ & N/A & 1.0 \\
    \hline
    \acs{DASSL} & N/A & $10^{-6}$ & N/A & Sim Time/500 \\
    \hline
    Proposed Tech. & Dynamic & $0$ & $10^{-3}$ & N/A \\
    \hline
  \end{tabular}
\end{table}

The configuration of the tools in listed in Table~\ref{tab:techs}. In
Table~\ref{tab:techs}, $\Delta q$ is the absolute quantum,
$\epsilon_{v}$ is the error tolerance of the values computed from
integrators. \ac{QSS}, RK-45 and \ac{DASSL} techniques also use the
$\epsilon_{v}$ in their level crossing detection algorithms. Parameter
$\epsilon_{t}$ is the error tolerance in the time domain, when finding
roots of polynomials, needed for the proposed technique. Finally,
maximum $\Delta t$ is the maximum integration step that fourth-fifth
order \ac{RK}-45 and \ac{DASSL} integration techniques can take. The
maximum $\Delta t$ value in Table~\ref{tab:techs} is the default value
used by the respective tools. In addition to the above set-up, the
proposed technique uses five terms in the Taylor polynomials when
approximating the transcendental functions.

We would like to point out that only \ac{QSS}-1 gave correct results in
Ptolemy, and hence we compare with only \ac{QSS}-1. Furthermore,
\ac{QSS} integration does not interact well with \ac{HA} semantics. The
advantage of \ac{QSS}, i.e., integrators running asynchronously to each
other, becomes a hindrance in correct execution of the \ac{HA}. Consider
the \ac{Robot} \ac{HA} in Figure~\ref{fig:robotha}. In order to detect
that the guard $g$ on the edge connecting $\mathbf{T1}$ and
$\mathbf{T2}$ is enabled, the value of continuous variables $x$ and $y$
should be available at the same time. However, due to the asynchronous
nature of the \ac{QSS} integrators these values are \textit{not}
available in the same time instant. Hence, the guard never triggers and
the level crossing is missed, resulting in incorrect behaviour. We
enforced synchronous execution of all the \ac{QSS} integrators by
triggering every integrator when any one integrator produces an output.

\subsection{Results}
\label{sec:results}

\begin{table}[tb]
  \centering
  \caption{Number of simulation steps taken by different techniques.
    \\MLC: Missed Level Crossing.}
  \label{tab:steptaken}
  \begin{tabular}{|c|p{30pt}|p{30pt}|p{30pt}|c|}
    \hline
    Benchmarks & Ptolemy-\ac{QSS}-1 & Ptolemy-\ac{RK}-45 & Open-Modelica-\ac{DASSL}
    & Proposed Tech. \\
    \hline
    TH & 3361 & 38 & 78 & 54 \\
    \hline
    WLM & 38000 & 61 & 252 & 19 \\
    \hline
    Robot & 1662 & 13 & MLC & 41 \\
    \hline
    AFb & 3170 & 41 & 151 & 36 \\
    \hline
    Reactor & 72503 & 68  & MLC  & 15 \\
    \hline
  \end{tabular}
\end{table}

We present two sets of results: \textcircled{1} execution time in terms
of the number of discrete simulation steps taken by each of the
simulation tool. \textcircled{2} Correctness of the proposed technique
by comparing the output traces.

\subsubsection{Execution time}
\label{sec:exec-time-speed}

Table~\ref{tab:steptaken} gives the number of steps taken by each of the
simulation tool. We compare in terms of simulation steps
like~\cite{kofman2004discrete}. One cannot directly compare the absolute
execution time, because each tool is implemented using different
programming languages and libraries, Ptolemy in Java, OpenModelica in
C++ and the proposed technique in Python.

Out of the five benchmarks, OpenModelica was able to simulate the least
number of benchmarks correctly. For our benchmarks, on average, the
proposed technique takes $\approx 720 \times$ fewer steps than
\ac{QSS}-1, $1.33\times$ fewer steps than \ac{RK}-45 solver and around
$4.41\times$ fewer steps than OpenModelica with the \ac{DASSL} solver.
The proposed technique and Ptolemy' level crossing detector were able to
handle equality on edge guards, for the \ac{Reactor} benchmark,
correctly. OpenModelica on the other hand resulted in missed level
crossing detections (indicated as MLC in Table~\ref{tab:steptaken}). The
proposed technique is correctly able to handle equality, because, unlike
other techniques, simulation always converges towards the first level
crossing and never overshoots it as shown in Theorems~\ref{thm:2}
and~\ref{thm:3}.

In our simulation framework the non-deterministic \ac{Reactor} \ac{HA}
is converted into a deterministic \ac{HA}, by \textit{always} choosing
the \textit{same} outgoing edge, from any given state, when more than
one outgoing edge are enabled. In case of Ptolemy one of the outgoing
edge is chosen randomly during program
execution~\cite{ptolemaeus2014system}.

\subsubsection{Correctness of the proposed technique}
\label{sec:corr-prop-techn}

\begin{table}[tb]
  \centering
  \caption{Correlation Coefficients. Ideal value is 1.0}
  \label{tab:correlation}
  \begin{tabular}{|c|c|c|p{35pt}|}
    \hline
    Benchmarks & Ptolemy-\ac{QSS}-1 & Ptolemy-\ac{RK}-45 & Open-Modelica-\ac{DASSL} \\
    \hline
    TH &  0.999986 & 1.0 & 1.0 \\
    \hline
    WLM &  1.0 & 0.865148 & 1.0 \\
    \hline
    Robot & 1.0 & 1.0 & N/A \\
    \hline
    AFb & 0.999861& 0.999409 & 0.999999\\
    \hline
    Reactor & 1.0 & 1.0 & N/A \\
    \hline
  \end{tabular}
\end{table}

Closed form solutions are not possible for all \acp{ODE} in our
benchmarks, e.g., the \acp{ODE} in the \ac{AFb} benchmark have no closed
form solutions. We compare the output trace from the proposed technique
with output traces of other techniques to validate the soundness of the
proposed approach. Given an output trace, for any benchmark, we obtain
the discrete time events, during simulation, when the level-crossing was
detected. We correlate these discrete time instants from the proposed
technique with \ac{QSS}-1, \ac{RK}-45 and \ac{DASSL} outputs. A
correlation coefficient of 1.0 indicates that the two techniques detect
level-crossings at the exact same discrete time instants.

Table~\ref{tab:correlation} gives the correlation coefficients relating
the discrete time instants, for level crossings, generated from the
proposed technique with other simulation techniques. As we can see,
\ac{QSS}-1, from Ptolemy, and \ac{DASSL} from OpenModelica, detect the
level-crossings at the same time as the proposed technique. However,
Ptolemy' \ac{RK}-45 integration technique combined with level-crossing
detection is less correlated with the proposed technique. \ac{RK}-45
integration takes fewer number of steps, than \ac{QSS}-1 and \ac{DASSL},
and hence, timing errors accumulate during simulation, which results in
under or over-estimating the discrete time points for level-crossings.
The output traces of the proposed technique and \ac{RK}-45 can be much
more closely related by reducing the maximum $\Delta t$. However, doing
so would increase the number of integration steps in \ac{RK}-45. The
proposed technique gives an ideal trade-off between the number of
integration/simulation steps vs.\ accuracy.


%% file: conclusion.tex
\section{Conclusion and Future Work}
\label{sec:conclusion}

\acf{HA} is a formal approach for specification and validation of safety
critical controllers. Simulation of \ac{HA} is challenging, because of
sudden discontinuities caused by level crossings that need to be
correctly detected. The current state-of-the-art level crossing
detection techniques can potentially miss detecting the level crossing
leading to incorrect behaviour.

In this work we propose a new formalism called \acf{QSHA} and an
associated discrete event simulation framework, which guarantees
semantics preserving program behaviour. The primary idea is to select
discrete simulation steps based on a \textit{dynamic} quantum in any
location of the \ac{QSHA}. The resultant simulation framework has been
proven to always make progress and converge to the first level crossing.
Interestingly the dynamic quantum based discrete event simulation
approach is quite efficient, outperforming \acf{QSS}-1, \acf{RK}-45, and
\acf{DASSL} based integration techniques combined with state-of-the-art
level crossing detectors.

The current approach syntactically composes a network of \acfp{HA} into
a single \ac{QSHA} before simulation. This approach is know to result in
state space explosion. In this future we plan to address this
shortcoming by developing a modular discrete event simulation framework,
which can simulate each \ac{HA} in the network individually.
